\documentclass[prl, twocolumn, superscriptaddress, footinbib, floatfix, amsmath, amsfonts, amssymb, amsthm]{revtex4-2}

\usepackage[utf8]{inputenc}
\usepackage[T1]{fontenc}
\usepackage[english]{babel}

\usepackage{graphicx}
\usepackage[caption=false]{subfig}

\usepackage{siunitx}
\usepackage[version=4]{mhchem}

\usepackage{hyperref}
\hypersetup{colorlinks=true, linkcolor=blue, citecolor=blue, urlcolor=blue, filecolor=red, pdfstartview=}

\newcommand*{\iu}{\mathrm{i}}
\newcommand*{\Elr}{\mathrm{e}}

\usepackage{upgreek}
\newcommand*{\Pauli}{\upsigma}

\usepackage{bm}
\newcommand*{\vdot}{\bm{\cdot}}
\newcommand*{\vcross}{\bm{\times}}
\newcommand*{\grad}{\bm{\nabla}}
\newcommand*{\vb}[1]{\bm{#1}}
\newcommand*{\vu}[1]{\bm{\hat{#1}}}

\newcommand*{\abs}[1]{\left\lvert {#1} \right\rvert}

\begin{document}
\title{Superconductivity due to fluctuating loop currents}
\author{Grgur Palle}
\email{grgur.palle@kit.edu}
\affiliation{Institute for Theoretical Condensed Matter Physics, Karlsruhe Institute of Technology, 76131 Karlsruhe, Germany}
\author{Risto Ojaj\"{a}rvi}
\affiliation{Institute for Theoretical Condensed Matter Physics, Karlsruhe Institute of Technology, 76131 Karlsruhe, Germany}
\author{Rafael M.\ Fernandes}
\affiliation{School of Physics and Astronomy, University of Minnesota, Minneapolis, Minnesota 55455, USA}
\author{J\"{o}rg Schmalian}
\affiliation{Institute for Theoretical Condensed Matter Physics, Karlsruhe Institute of Technology, 76131 Karlsruhe, Germany}
\affiliation{Institute for Quantum Materials and Technologies, Karlsruhe Institute of Technology, 76131 Karlsruhe, Germany}
\date{\today}
\begin{abstract}
Orbital magnetism and the loop currents (LC) that accompany it have been proposed to emerge in many systems, including cuprates, iridates, and kagome superconductors. In the case of cuprates, LCs have been put forward as the driving force behind the pseudogap, strange-metal behavior, and $d_{x^2-y^2}$-wave superconductivity. Here, we investigate whether fluctuating intra-unit-cell loop currents can cause unconventional superconductivity.
For odd-parity LCs, we find that they are strongly repulsive in all pairing channels near the underlying quantum-critical point (QCP).
For even-parity LCs, their fluctuations do give rise to unconventional pairing. However, this pairing is not amplified in the vicinity of the QCP, in sharp contrast to other known cases of pairing mediated by intra-unit-cell order parameters, such as spin-magnetic, nematic, or ferroelectric ones.
Applying our formalism to the cuprates, we conclude that pairing mediated by fluctuating intra-unit-cell LCs is unlikely to yield $d_{x^2-y^2}$-wave superconductivity.
We also show that loop currents, if relevant for the cuprates, must vary between unit cells and break translation symmetry.
\end{abstract}

\maketitle

\section*{Introduction}
Although magnetic order most commonly arises from interactions related to the spin degrees of freedom, in correlated systems magnetism may also develop in the orbital sector. Whenever such orbital magnetism occurs, time-reversal symmetry breaking manifests itself through a pattern of spontaneously flowing currents.
This pattern must be made of closed loops to avoid a global current, forbidden due to a theorem by Bloch~\cite{Bohm1949,Ohashi1996,Watanabe2022}.
Through the years, many types of loop-current (LC) patterns have been proposed in a variety of systems.
In the cuprates, inversion-symmetry-breaking intra-unit-cell LCs have been put forward as the underlying order of the pseudogap state~\cite{Varma1997,Simon2003,Varma2006}, while their fluctuations have been proposed to drive both $d_{x^{2}-y^{2}}$-wave superconductivity~\cite{Aji2010} and marginal Fermi liquid behavior near the quantum-critical point (QCP)~\cite{Aji2007,Varma1987}.
Cuprate LC order was also invoked to explain polarized neutron scattering experiments~\cite{Fauque2006, Bourges2008, Greven2008,Greven2010, Bourges2021}, although alternative interpretations exist~\cite{Croft2017}.
A state consistent with LC order has been inferred from second-harmonic generation measurements in the iridate Mott insulator \ce{Sr2IrO4}~\cite{Zhao2016} which displays an unusual gap upon doping~\cite{Yan2015,Kim2016}.
A LC pattern that breaks translation symmetry is one of the main candidates for explaining why the charge-density wave displayed by the recently discovered kagome superconductors seemingly breaks time-reversal symmetry~\cite{Mielke2022}.
Beyond specific materials, LCs have also been discussed in the context of the spontaneous anomalous Hall effect in Fermi liquids~\cite{Sun2008, Castro2011, Sur2018} and in the context of spin liquids with broken time-reversal symmetry~\cite{Scheurer2018}.

\begin{figure}
\centering
\includegraphics[width=0.85\columnwidth]{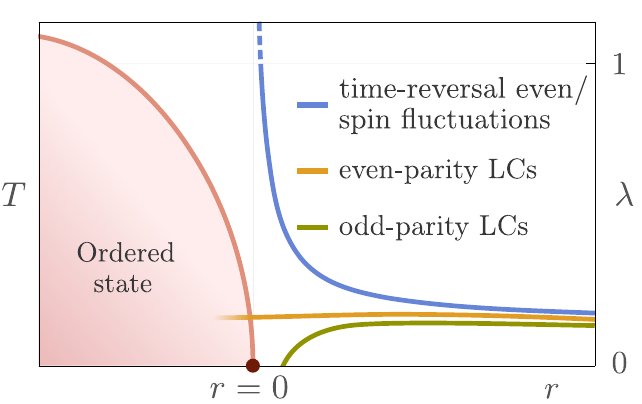}
\caption{Schematic behavior of the superconducting transition temperature and the leading pairing eigenvalue $\lambda$ near a QCP controlled by the tuning parameter $r$.
The superconducting transition temperature is given by $T_{c} \propto \omega_c \Elr^{-1/\lambda}$. Pairing mediated by time-reversal-even charge-fluctuations or spin-fluctuations (blue) is enhanced near the QCP, where weak-coupling theory breaks down (dashed line).
In contrast, we find that the pairing mediated by even-parity loop currents (yellow) is not enhanced at the QCP, whereas pairing mediated by odd-parity loop currents (green) becomes strongly repulsive near the QCP.}
\label{fig:QCP}
\end{figure}

Given their potential realization in a diverse set of systems, it is important to elucidate whether fluctuating loop currents can give rise to superconductivity.
In this context, intra-unit-cell (i.e., $\vb{q} = \vb{0}$) LCs have been prominently discussed as the pairing glue of the cuprates which makes them especially interesting, notwithstanding the difficulties in detecting them.
For comparison, in the case of fluctuations from intra-unit-cell orders that preserve time-reversal symmetry, such as nematic~\cite{Lederer2015,Lederer2017,Klein2018} and ferroelectric~\cite{Kozii2015,Klein2023} ones, it is well established that $s$-wave pairing generally emerges with a number of attractive subleading channels.
Moreover, superconductivity in these cases is strongly enhanced as the QCP is approached, thus establishing a robust regime in which pairing is dominated by the corresponding fluctuations.
Pairing is promoted by ferromagnetic spin fluctuations~\cite{Millis2001,Chubukov2003} as well, the main difference being the $p$-wave nature of the leading pairing state.
However, the case of pure orbital magnetism is different, not only because LCs do not directly couple to the spin, but also because they usually break additional symmetries besides time reversal.
This leads to two important questions of broad and particular relevance.
First, are there general conditions, independent of the details of a given material, under which pairing is dominated by quantum-critical intra-unit-cell LC fluctuations?
Second, in the specific case of the cuprates, can fluctuating intra-unit-cell LCs cause or enhance $d_{x^{2}-y^{2}}$-wave pairing?

In this paper, we answer both questions.
We show that LC fluctuations do not give rise to an enhanced pairing near the QCP, as shown schematically in Fig.~\ref{fig:QCP}.
Even-parity LCs, such as orbital ferromagnets or orbital altermagnets, may cause unconventional pairing.
However, they are as likely or unlikely to do so as any other degree of freedom far from its critical point. This is because the pairing promoted by these fluctuations is not enhanced as the QCP is approached (yellow line in Fig.~\ref{fig:QCP}), in sharp contrast to the cases of ferromagnetic spin fluctuations or time-reversal-even charge fluctuations, such as nematic or ferroelectric ones (blue line in Fig.~\ref{fig:QCP}).
LCs that break parity, i.e.\ states of magneto-electric order, are repulsive for all pairing symmetries as one approaches the QCP (green line in Fig.~\ref{fig:QCP}).
Hence they weaken pairing caused by other mechanisms.
Such odd-parity LC states can at best support superconductivity when their fluctuations are sufficiently weak.
In the context of the cuprates, we show that among the three candidate LC states, shown in Fig.~\ref{fig:CuO2_plane}(b)--(d), only the fluctuations of the parity-preserving $d$-wave LC favor weak $d_{x^{2}-y^{2}}$-wave pairing; see Fig.~\ref{fig:orbital-all}.
In the presence of weak spin-orbit coupling, triplet pairing mediated by secondary spin-magnetic fluctuations takes place (Fig.~\ref{fig:spin-all}) and always prevails for $d$-wave LCs as the QCP is approached.

\section*{Results}
\subsection*{Formalism}
We start with a general analysis that allows us to draw conclusions that are independent of material details.
Consider a centrosymmetric system with $M$ orbitals per primitive unit cell and introduce the spinors $c_{\vb{k}\sigma} = \left(c_{\vb{k} \sigma 1}, \ldots, c_{\vb{k} \sigma M}\right)^{\intercal}$ and $c_{\vb{k}} = \left(c_{\vb{k}\uparrow}, c_{\vb{k}\downarrow}\right)^{\intercal}$ in terms of which the one-particle Hamiltonian equals $\mathcal{H}_0 = \sum_{\vb{k}} c_{\vb{k}}^{\dag} H_{\vb{k}} c_{\vb{k}}$.
For $\mathcal{H}_{0}$, we assume that it preserves parity and time-reversal symmetry.
We treat the interactions phenomenologically and assume from the outset that they give rise to intra-unit-cell orbital magnetism and LCs.

Under these assumptions, the interacting Hamiltonian $\mathcal{H}_{\mathrm{int}} = g \sum_{\vb{q}} \Phi_{-\vb{q}} \phi_{\vb{q}}$ can be described in terms of a coupling between the fluctuating LC order parameter $\Phi_{\vb{q}}$ and a symmetry-appropriate fermionic bilinear
\begin{equation}
\phi_{\vb{q}} = \frac{1}{\sqrt{N}} \sum_{\vb{k}} c_{\vb{k}}^{\dag} \Gamma_{\vb{k},\vb{k}+\vb{q}} c_{\vb{k}+\vb{q}}. \label{eq:phi}
\end{equation}
Here $N$ is the number of unit cells and $g$ is the coupling constant.
The orbital LC pattern associated with $\Phi_{\vb{q}}$ is encoded in the form factor $\Gamma_{\vb{k},\vb{p}} = \Gamma_{\vb{p},\vb{k}}^{\dag}$, which is a matrix in spin and orbital space.
In the absence of spin-orbit coupling, these form factors are trivial in spin space, meaning $\Gamma_{\vb{k},\vb{p}} = \gamma_{\vb{k},\vb{p}} \otimes \Pauli^{0}$ where $\Pauli^{0}$ is the identity matrix in spin space.
Consequently, the orbital matrix must be odd under time reversal, $\gamma_{\vb{k},\vb{p}}^{*} = - \gamma_{-\vb{k},-\vb{p}}$.

LC fluctuations are described by the $\Phi_{\vb{q}}$ correlation function $\chi(\vb{q},\omega)$, which we assume to be peaked at $\vb{q} = \vb{0}$ in momentum space and characterized by a correlation length $\xi = a_0 r^{-\nu}$, where $a_0$ is a microscopic length scale.
$r$ is a dimensionless parameter that, by definition, vanishes at the QCP and is $r \sim 1$ for a structureless correlation function in momentum space.
For the static correlation function we use the critical scaling expression $\chi(\vb{q}) = \mathcal{F}(q\xi)/q^{2-\eta}$ with critical exponents $\nu$ and $\eta$, and scaling function $\mathcal{F}(y)$ that has the usual asymptotic behaviors $\mathcal{F}(y \gg 1) \sim \mathrm{const.}$ and $\mathcal{F}(y \ll 1) \sim y^{2-\eta}$.

\begin{figure}[t]
\centering
\includegraphics[width=\columnwidth]{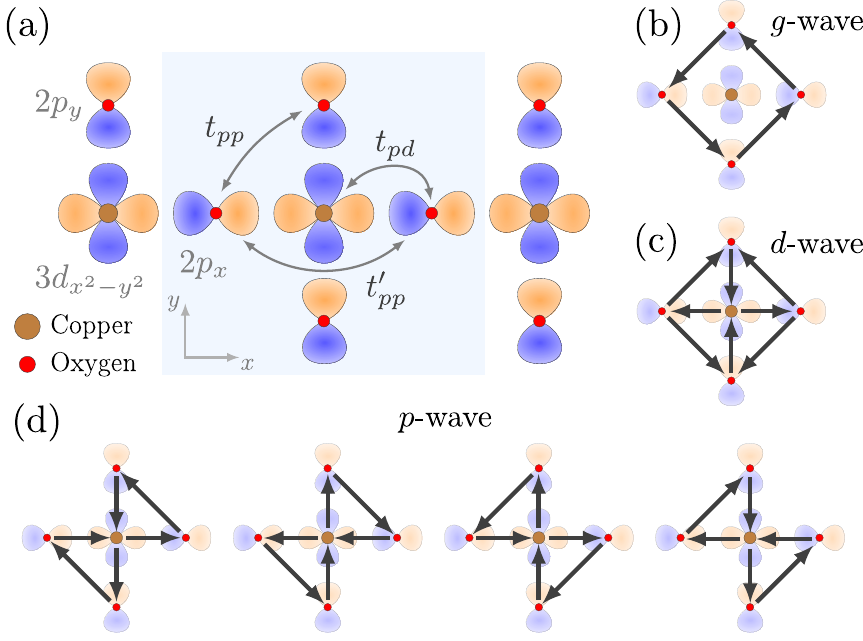}
\caption{(a) The \ce{CuO2} plane of the cuprates and its active orbitals. Arrows denote hoppings included in our model. Light blue shading highlights the five orbitals used to form loop currents. (b)--(d) The three possible LC patterns of the \ce{CuO2} plane with (b) $g_{xy(x^2-y^2)}$, (c) $d_{x^2-y^2}$, and (d) $(p_x, p_y)$ character. While the last one has the symmetry of a four-state clock-model, as indicated by the four degenerate patterns, the former two display Ising symmetry.}
\label{fig:CuO2_plane}
\end{figure}

Solving this coupled many-body problem is a formidable challenge.
In order to make progress, we follow the strategy of Refs.~\cite{Aji2010,Lederer2015} and consider the system in the regime where the coupling of the electrons to LC fluctuations is sufficiently weak ($g \to 0$).
This allows us to analyze the pairing instability to leading-order in perturbation theory.
Such a strategy should be reasonable on the disordered side, far enough from the QCP, 
where Fermi liquid behavior is established and collective fluctuations are
sufficiently weak.
Importantly, this approach enables us to determine whether or not the weak-coupling theory breaks down as one approaches the QCP, thus providing an indicator for strong quantum-critical pairing.

In the weak-coupling limit, it is straightforward to derive the linearized gap equation for the singlet and triplet pairing channels (Appendix~\ref{sec:LC-exchange}):
\begin{equation}
\int_{\mathrm{FS}} \frac{dS_{\vb{k}}}{v_{\vb{k}}} V_{s}(\vb{p},\vb{k}) \Delta_{s}(\vb{k}) = \lambda \, \Delta_{s}(\vb{p}). \label{eq:gapeq}
\end{equation}
Here the integral goes over the Fermi surface, $v_{\vb{k}} = \abs{\grad \varepsilon_{\vb{k}}}$ is the Fermi velocity, and $s = +1$ ($-1$) stands for singlet (triplet) pairing.
The largest eigenvalue $\lambda$ determines the superconducting transition temperature through $k_{B} T_{c} = \omega_{c} \tfrac{2\Elr^{\gamma}}{\pi} \Elr^{-1/\lambda}$, where $\omega_{c}$ is the characteristic cutoff for LC fluctuations and $\gamma$ is the Euler–Mascheroni constant.
The eigenvector $\Delta_{s}(\vb{p})$ determines the symmetry of the pairing and is related to the superconducting gap function of the Bogoliubov-de~Gennes Hamiltonian via
$\Delta_{\sigma\sigma'}(\vb{p}) = \Delta_{+}(\vb{p}) \iu \Pauli^{y}_{\sigma\sigma'}$ for singlet and
$\Delta_{\sigma\sigma'}(\vb{p}) = \Delta_{-}(\vb{p}) \left(\Pauli^{a} \iu \Pauli^{y}\right)_{\sigma\sigma'}$ for triplet pairing, respectively.
Here all triplet orientations are degenerate because we assumed no spin-orbit coupling and purely orbital LCs $\Gamma_{\vb{k},\vb{p}} = \gamma_{\vb{k},\vb{p}} \otimes \Pauli^{0}$.

The Cooper channel interactions are given by
\begin{equation}
V_{\pm}(\vb{p},\vb{k}) = - \frac{1}{2}\big(V_0(\vb{p},\vb{k}) \pm V_0(\vb{p},-\vb{k})\big), \label{eq:pairing_int}
\end{equation}
where the overall minus sign arises because LCs are odd under time reversal.
$V_0(\vb{p},\vb{k}) = g^{2} \chi(\vb{p}-\vb{k}) f(\vb{p},\vb{k})$ is a combination of the LC correlation function $\chi(\vb{q})$ and the matrix element
\begin{equation}
f(\vb{p},\vb{k}) = \abs{u_{\vb{p}}^{\dag} \gamma_{\vb{p},\vb{k}} u_{\vb{k}}}^2 > 0, \label{eq:aux_pairing_int}
\end{equation}
where $u_{\vb{k}}$ are the orbital components of the conduction band eigenvectors.
The $f(\vb{p},\vb{k})$ contain information about the nature and symmetry of the LC state via the form factor $\gamma_{\vb{p},\vb{k}}$.
For the coupling constant $g$, we assume a value that yields sufficiently small dimensionless eigenvalues $\lambda$.

\subsection*{Generic behavior near the quantum-critical point}
Before we discuss our results near the QCP, let us briefly comment on pairing mediated by critical fluctuations of other order parameters.
For order parameters that are spin-ferromagnetic or preserve time-reversal symmetry, as the QCP is approached ($r \to 0$) one finds that the largest eigenvalue of the gap equation diverges like $\lambda \propto r^{-\psi}$ with $\psi>0$~\cite{Abanov2001, Varma2012,Lederer2015}, as schematically shown by the blue line in Fig.~\ref{fig:QCP}.
While this corresponds to a breakdown of the weak-coupling analysis, it also signals the emergence of a strong pairing tendency near the QCP.
Weak-coupling theory alone cannot determine the precise behavior in the immediate vicinity of the QCP, yet numerous computational approaches show that $T_c$ is largest at or near the QCP~\cite{Wang2017,Xu2017,Xu2020}.
This is the much-discussed efficiency of quantum-critical pairing~\cite{ChubukovI_2020,ChubukovII_2020}.
Following Ref.~\cite{Lederer2015}, the  divergence of $\lambda$ is based on the assumption that the forward-scattering contribution $\left.f(\vb{p},\vb{k})\right|_{\vb{p} \to \vb{k}}$ is attractive and varies smoothly as a function of $\vb{q} = \vb{p}-\vb{k}$.
Under these circumstances, the largest eigenvalue of the gap equation is given by
\begin{equation}
\lambda \approx \lambda_0 \int d^{d-1}q_{\parallel} \, \chi(\vb{q}_{\parallel}), \label{lambda_qcp}
\end{equation}
where $\vb{q}_{\parallel}$ are the components of the transferred momentum $\vb{q}$ tangential to the Fermi surface and $\lambda_0 = g^{2} \left\langle v_{\vb{k}}^{-1} f(\vb{k},\vb{k})\right\rangle_{\mathrm{FS}}$.
Using the scaling form $\chi(\vb{q}_{\parallel}) = \mathcal{F}(q_{\parallel}\xi)/q_{\parallel}^{2-\eta}$ introduced previously, the integral in Eq.~\eqref{lambda_qcp} gives $\psi = (3-d-\eta)\nu$ if $d < 3-\eta$.
Hence QCPs in $d=2$ with $\eta<1$ yield strong pairing.
In $d=3$ the enhancement is logarithmic, provided $\eta=0$.

For intra-unit-cell orbital magnetism, however, the analysis of the pairing enhancement at the QCP is different and depends on the parity $p_{\Phi}$ of the LC state.
In Appendix~\ref{sec:LC-exchange}, we show that:
\begin{equation}
\begin{aligned}
\left.f(\vb{p},\vb{k})\right|_{\vb{p} \to \vb{k}} &\propto \left(\vb{p}-\vb{k}\right)^{2} &\text{if $p_{\Phi} = +1$,} \\
\left.f(\vb{p},\vb{k})\right|_{\vb{p} \to\vb{k}} &\propto \mathrm{const.} > 0 &\text{if $p_{\Phi} = -1$,}
\end{aligned} \label{eq:nearQCP}
\end{equation}
as well as $\left.f(\vb{p},\vb{k})\right|_{\vb{p} \to -\vb{k}} \propto \left(\vb{p}+\vb{k}\right)^{2}$ for both values of $p_{\Phi}$.
Hence, for even-parity LCs ($p_{\Phi}=+1$), the forward-scattering singularity as $r \to 0$ originating from $\chi(\vb{p}-\vb{k})$ in Eq.~\eqref{lambda_qcp} is eliminated for $d \geq 1-\eta$ and suppressed for $d < 1-\eta$ down to $\lambda \propto r^{(d-1+\eta)\nu}$.
Thus for $d \geq 1-\eta$, which we expect to always be fulfilled for systems of interest, the pairing response is not enhanced near the QCP, as illustrated in Fig.~\ref{fig:QCP} (yellow line).
For odd-parity LC ($p_{\Phi}=-1$), the implications of Eq.~\eqref{eq:nearQCP} are even more dramatic.
The positive-definiteness of the matrix element $f(\vb{p},\vb{k}) > 0$ when combined with the monotonous decay of $\chi(\vb{q})$ from its $\vb{q} = \vb{0}$ maximum imply that that all pairing channels are repulsive near the QCP, diverging as $\lambda \propto - r^{-\psi}$ (shown schematically by the green line in Fig.~\ref{fig:QCP}).
Only away from the QCP can finite-$\vb{q}$ features of the matrix element $f(\vb{p},\vb{k})$ or the Fermi velocity $v_{\vb{k}}$ result in an attractive pairing channel that, however, is parametrically weak.
This robust result does not depend on material details and is a consequence of the fact that LC order breaks time-reversal symmetry with a trivial form factor in spin space. 
Hence, for two-dimensional systems, even-parity intra-unit-cell LCs are inefficient and odd-parity intra-unit-cell LCs are detrimental to pairing near their QCP.

\begin{figure*}[t]
\centering
\includegraphics[width=\textwidth]{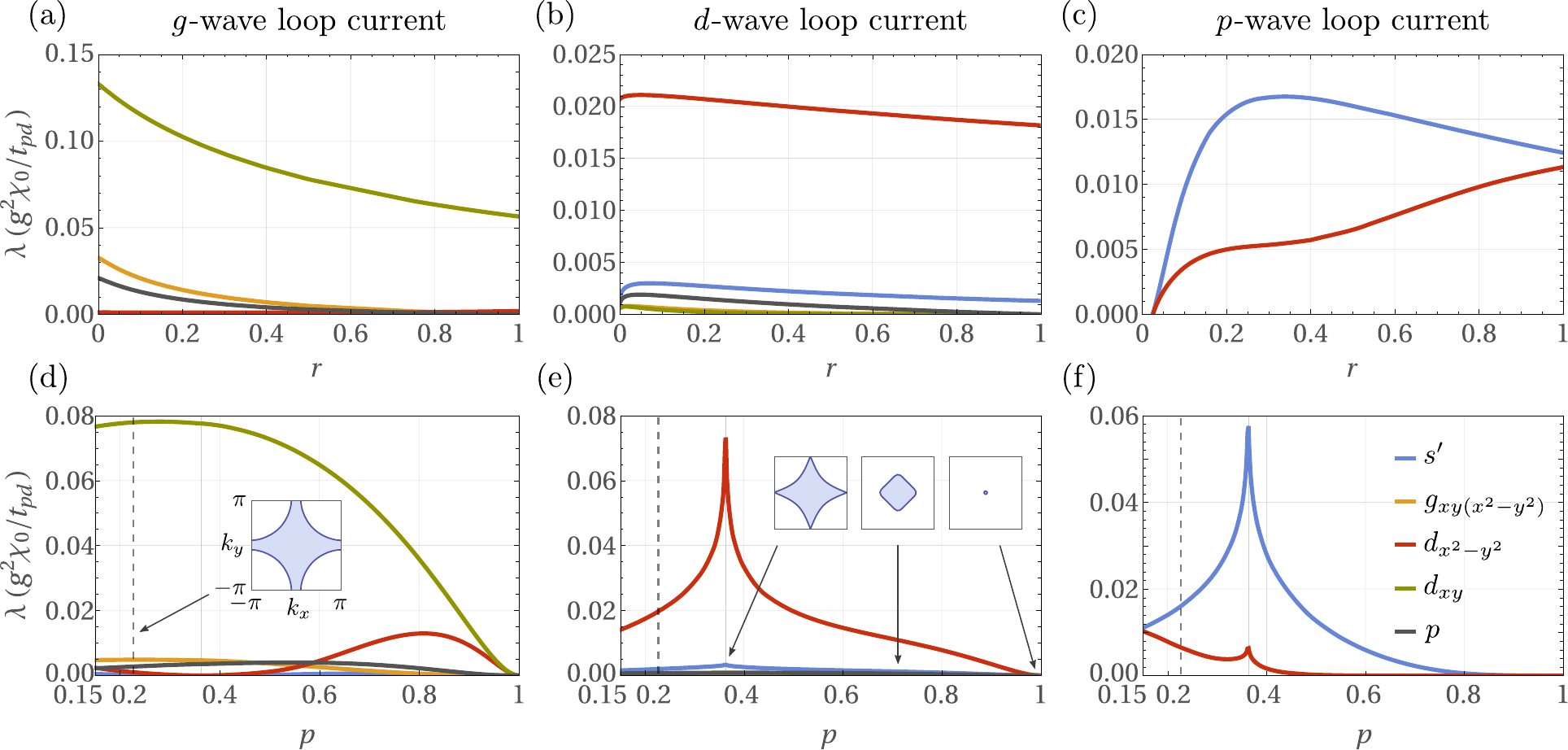}
\caption{Pairing eigenvalues $\lambda$ [Eq.~\eqref{eq:gapeq}] for the three LC states of Fig.~\ref{fig:CuO2_plane} as a function of the parameter $r$ characterizing the proximity to the QCP at fixed chemical potential $\mu = 0.9 t_{pd}$ [panels (a)--(c)]. Panels (d)--(f) show the eigenvalues at fixed $r = 0.5$ as a function of the hole doping $p$.
The dashed vertical lines in panels (d)--(f) denote the hole doping $p=0.23$ corresponding to $\mu = 0.9 t_{pd}$.
The insets in panels (d) and (e) illustrate the Fermi surfaces at different values of $p$. The Lifshitz transition occurs at $p=0.36$.
$\Phi_g$ [panels (a),(d)] and $\Phi_d$ [panels (b),(e)] fluctuations yield $d_{xy}$ and $d_{x^2-y^2}$ pairing, respectively, that is only weakly enhanced near the QCP.
$\vb{\Phi}_{p}$ fluctuations [panel (c)] yield extended s-wave pairing (denoted $s'$) at not too small $r$, turning repulsive as the QCP is approached ($r \to 0$).
There is a one-parameter family of possible $\vb{\Phi}_{p}$ parameterized by $\alpha$ (Fig.~\ref{fig:Eu-orbital-alpha}). In panels (c),(f), we use $\alpha = 0$.}
\label{fig:orbital-all}
\end{figure*}

\subsection*{Application to overdoped cuprates}
The conclusions drawn so far are valid for generic systems.
Now we consider the cuprates. We analyze them from the far-overdoped side of the phase diagram where the normal state is a Fermi liquid~\cite{Proust2002, Nakamae2003, Hussey2003, Damascelli2003, Plate2005, Horio2018, Vignolle2008, Bangura2010}. Provided the pairing state and dominant mechanism are unchanged across the phase diagram, this should give information about optimally doped systems as well. 
For the one-particle Hamiltonian $\mathcal{H}_0$, we employ the well-established three-band tight-binding model~\cite{Varma1987, Emery1987, Emery1988, Littlewood1989, Gaididei1988, Scalettar1991} that is based on the copper $3d_{x^{2}-y^{2}}$ and oxygen $2p_{x,y}$ orbitals; see Fig.~\ref{fig:CuO2_plane}(a) and Appendix~\ref{sec:CuO2-TBA-model}. This model is characterized by the charge-transfer energy $\epsilon_d - \epsilon_p$ and the hopping amplitudes shown in Fig.~\ref{fig:CuO2_plane}(a). We use the \ce{Cu}-\ce{O} hopping element $t_{pd} \approx \SI{1.4}{\electronvolt}$ to set the overall energy scale.
Although we considered a wide range of tight-binding parameters discussed in the literature~\cite{Pickett1989, Pavarini2001, Kent2008, Weber2014, Photopoulos2019}, the precise choice of tight-binding parameters has proven to have a minimal impact on our results.
We thus use one representative choice of parameters (Appendix~\ref{sec:CuO2-TBA-model}) throughout.

Interactions in the cuprates are most often modeled with extended Hubbard interactions.
Whether LC order emerges in the resulting model is under debate, as there are computational investigations that do~\cite{Weber2009,Weber2014,Tazai2021} and do not~\cite{Greiter2007,Thomale2008,Kung2014} find evidence for LCs.
While these are important microscopic investigations, we take a more phenomenological perspective and assume from the outset that intra-unit-cell LC fluctuations exist and exploit the consequences of this assumption.
In this phenomenological approach, we can independently vary the LC correlation length through $r$ and the hole doping $p$ through the chemical potential $\mu$. In the real system the two are not independent, something we must keep in mind when interpreting our results.
For the LC correlation function we use $\chi(\vb{q}) = \chi_{0} \left(\frac{1+r}{2} - \frac{1-r}{4} \nu_{\vb{q}}\right)^{-1}$ where $\chi_0 > 0$ and $\nu_{\vb{q}} = \cos q_{x} + \cos q_{y}$; negative $\chi(\vb{q})$ indicates LC condensation.
As we do not know in which LC channel the system orders, we classify all the possibilities (Fig.~\ref{fig:CuO2_plane}(b)--(d), Appendix~\ref{sec:bilinears}) and study Cooper pairing for each.
Figure~\ref{fig:orbital-all} shows the results.

To classify intra-unit-cell LCs, we consider the minimal set of sites that maps onto itself under all point group operations, namely, the set made of one \ce{Cu} site and the four surrounding \ce{O} sites; see the blue-shaded region of Fig.~\ref{fig:CuO2_plane}(a).
The corresponding five-component spinor $a_{\vb{k}\sigma} = \left(d_{\vb{k}\sigma}, p_{x,\vb{k}\sigma}, p_{y,\vb{k}\sigma}, \Elr^{-\iu k_x} p_{x,\vb{k}\sigma}, \Elr^{-\iu k_y} p_{y,\vb{k}\sigma}\right)^{\intercal}$ has particularly simple symmetry transformation rules, facilitating the group-theoretic classification (Appendix~\ref{sec:bilinears}).
In total, there are $25$ Hermitian matrices $\Lambda$ that one may use to construct an orbital intra-unit-cell fermionic bilinear $\phi(\vb{R}) = \sum_{\sigma} a_{\sigma}^{\dag}(\vb{R}) \Lambda a_{\sigma}(\vb{R})$.
By projecting onto the Bloch states via $a_{\vb{k}\sigma} = W_{\vb{k}} c_{\vb{k}\sigma}$ where $c_{\vb{k}\sigma} = \left(d_{\vb{k}\sigma}, p_{x,\vb{k}\sigma}, p_{y,\vb{k}\sigma}\right)^{\intercal}$, one finds the form factors $\gamma_{\vb{k},\vb{p}} = W_{\vb{k}}^{\dag}\Lambda W_{\vb{p}}$ of Eq.~\eqref{eq:phi}.
Since $W_{\vb{k}}^{*} = W_{-\vb{k}}$, the condition $\gamma_{\vb{k},\vb{p}}^{*} = - \gamma_{-\vb{k},-\vb{p}}$ implies that LCs have purely imaginary $\Lambda$.
The purely imaginary nature of the orbital matrix $\Lambda$ can be interpreted as introducing phase shifts in the bare hopping parameters of $\mathcal{H}_0$.
Via a reverse Peierls substitution, these phase shifts correspond to magnetic fluxes generated by orbital currents.

\begin{figure}[t]
\centering
\includegraphics[width=\columnwidth]{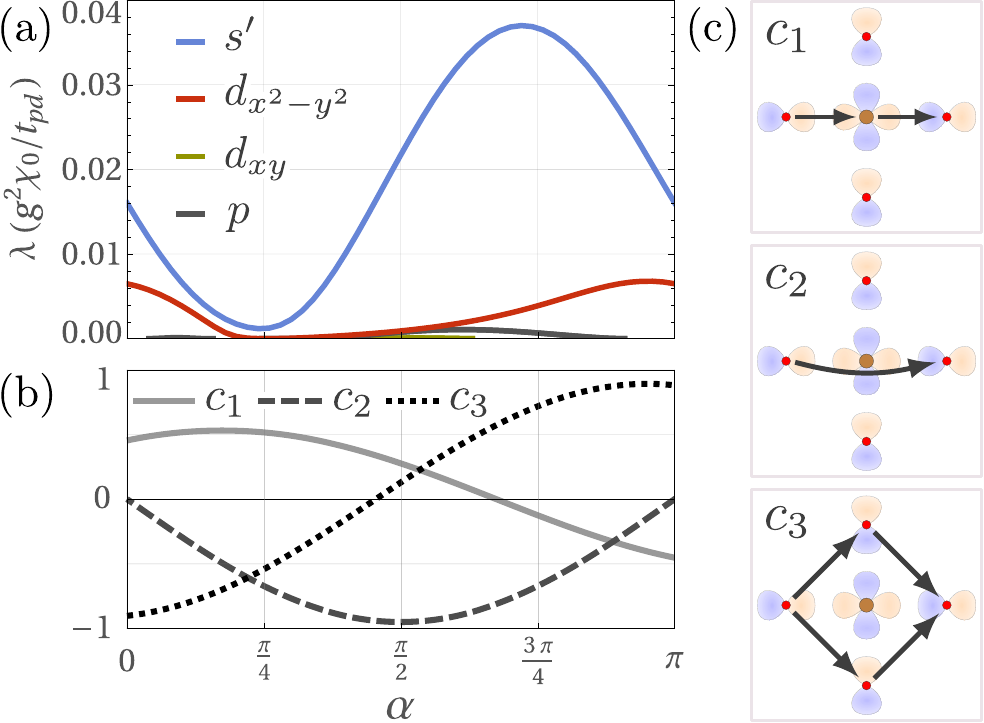}
\caption{(a) Pairing eigenvalues $\lambda$ [Eq.~\eqref{eq:gapeq}] due to $p$-wave LC fluctuations as a function of $\alpha$ for fixed $r=0.5$ and $\mu = 0.90 t_{pd}$.
$s'$ refers to an extended $s$-wave state dominated by $\cos(4 \varphi)$ dependence on the Fermi surface angle $\varphi$.
(b) $\alpha$ parametrizes the coefficients of the three $p$-wave current components $c_{1,2,3}$ illustrated in panel (c).
The coefficients $c_{1,2,3}$ are constrained to not generate a global current (Appendix~\ref{sec:current-constraints}).}
\label{fig:Eu-orbital-alpha}
\end{figure}

There are in total $10 = \frac{5(5-1)}{2}$ imaginary Hermitian matrices $\Lambda$.
We chose them so that they transform under irreducible representations (irreps) of the tetragonal point group $D_{4h}$~\cite{Dresselhaus2007}.
These irreps, in turn, determine which additional crystallographic symmetries are broken (if any) by the LC, besides time reversal.
The explicit expressions of the LC $\Lambda$ matrices are provided in Appendix~\ref{sec:bilinears}.
Upon enforcing the constraints that no global currents are allowed~\cite{Bohm1949,Ohashi1996,Watanabe2022} and that the currents obey Kirchhoff's law at steady state, we are left with a total of six LC patterns; see Appendix~\ref{sec:current-constraints}.
Of these six LC $\Lambda$, one is $g_{xy(x^2-y^2)}$-wave, one is $d_{x^2-y^2}$-wave, and two pairs are $(p_x,p_y)$-wave; see Fig.~\ref{fig:CuO2_plane}(b)--(d).
The corresponding order parameters we shall call $\Phi_g$, $\Phi_d$, and $\vb{\Phi}_{p} = (\Phi_{p_{x}},\Phi_{p_{y}})$.
As shown in Fig.~\ref{fig:Eu-orbital-alpha}, one may interpolate between the two $p$-wave options, which we parametrize with $\alpha \in [0, \pi]$.
This follows from the existence of several paths connecting opposite oxygen orbitals of the same kind: an indirect path through the \ce{Cu} atom (process $c_1$ in Fig.~\ref{fig:Eu-orbital-alpha}(c)), a direct path (process $c_2$), and an indirect path through the \ce{O} atoms (process $c_3$).
In the actual cuprate structure, the second process is mediated by the \ce{Cu}:$4s$ orbital~\cite{Andersen1995, Pavarini2001, Kent2008}.
In Figs.~\ref{fig:CuO2_plane}(d) and \ref{fig:orbital-all} we use $\alpha = 0$.
These LCs are essentially the same ones that were discussed in Ref.~\cite{Aji2010}, with the exception of one LC state discussed therein that breaks translation invariance.

Using the form factors $\Gamma_{\vb{k},\vb{p}} = W_{\vb{k}}^{\dag}\Lambda W_{\vb{p}} \otimes \Pauli^0$ from the analysis above, we numerically solve the linearized gap equation~\eqref{eq:gapeq} supplemented by Eqs.~\eqref{eq:pairing_int} and~\eqref{eq:aux_pairing_int}.
We consider two tuning parameters: the distance to the QCP $r$ and the chemical potential $\mu$,. The latter determines the hole doping concentration $p$ and the shape of the Fermi surface, which crosses the van Hove singularity (VHS) at $p=0.36$; see inset of Fig.~\ref{fig:orbital-all}(d),(e).
All eigenvalues $\lambda$ are measured in terms of the dimensionless parameter $g^2 \chi_0 / t_{pd}$ and are thus comparable.
The results are shown in Fig.~\ref{fig:orbital-all}.

$\Phi_{g}$ describes a $g_{xy(x^2-y^2)}$-wave LC which gives rise to an orbital-magnetic dipole, i.e., an orbital ferromagnet.
It has even parity ($p_{\Phi}=+1$) and transforms under the $A_{2g}$ irrep of $D_{4h}$.
$\Phi_{g}$ is an Ising order parameter and can be polarized by an external magnetic field orientated along the $z$ direction $B_{z}$ via the coupling $\mathcal{H}_{c} = - \kappa \Phi_{g} B_{z}$, where $\kappa$ is a coupling constant.
As shown in Fig.~\ref{fig:orbital-all}(a), $\Phi_{g}$ fluctuations result in weak $d_{xy}$ pairing, which is weak in the sense that the pairing eigenvalue $\lambda$ does not diverge at the QCP ($r \to 0$).
This is in agreement with Eq.~\eqref{eq:nearQCP} and the general result discussed thereafter.
There are sub-leading singlet and triplet instabilities as well.
In Fig.~\ref{fig:orbital-all}(d) one sees that the leading $d_{xy}$ instability is weakly enhanced near the VHS, while $d_{x^2-y^2}$ pairing is strongly suppressed in the same limit.
The reported~\cite{Aji2010} degeneracy between $d_{xy}$ and $d_{x^2-y^2}$ pairing for $\Phi_g$ is recovered in the limit of extremely overdoped systems with small Fermi surfaces, $p \to 1$.
The counter-intuitive result that this degeneracy is lifted in favor of $d_{xy}$ pairing by realistic $\mu$ values follows from the fact that the matrix element $f(\vb{p},\vb{k})$ vanishes whenever either $\vb{p}$ or $\vb{k}$ are at the high-symmetry Van Hove points $(\pm\pi,0)$ or $(0,\pm\pi)$, as proved in Appendix~\ref{sec:bilinears}.
Hence $\Phi_{g}$-mediated pairing cannot exploit the enhanced density of states due to the VHS.

The order parameter $\Phi_{d}$ is associated with $d_{x^{2}-y^{2}}$-wave LCs and is a magnetic octupole, i.e., an orbital altermagnet that is invariant under the combination of time reversal and a four-fold rotation about the $z$ axis, $\Theta C_{4z} \Phi_{d} = \Phi_{d}$.
It transforms under the $B_{1g}$ irrep and as such it has even parity, $p_{\Phi}=+1$.
Like $\Phi_{g}$, $\Phi_{d}$ is an Ising order parameter, but unlike $\Phi_{g}$, it does not have a magnetic moment.
Instead, it displays piezomagnetism and can be polarized by the combination of shear strain $\epsilon_{xy}$ and an external magnetic field pointing in the $z$ direction: $\mathcal{H}_{c} = - \kappa \Phi_{d} B_{z} \epsilon_{xy}$.
As shown in Fig.~\ref{fig:orbital-all}(b), $\Phi_{d}$ promotes weak $d_{x^{2}-y^{2}}$ pairing with several sub-leading singlet and one triplet pairing instabilities.
The pairing strength of the leading $d_{x^{2}-y^{2}}$  channel is enhanced if one tunes the chemical potential to the VHS, as can be seen in Fig.~\ref{fig:orbital-all}(e).

\begin{figure*}[t]
\centering
\includegraphics[width=\textwidth]{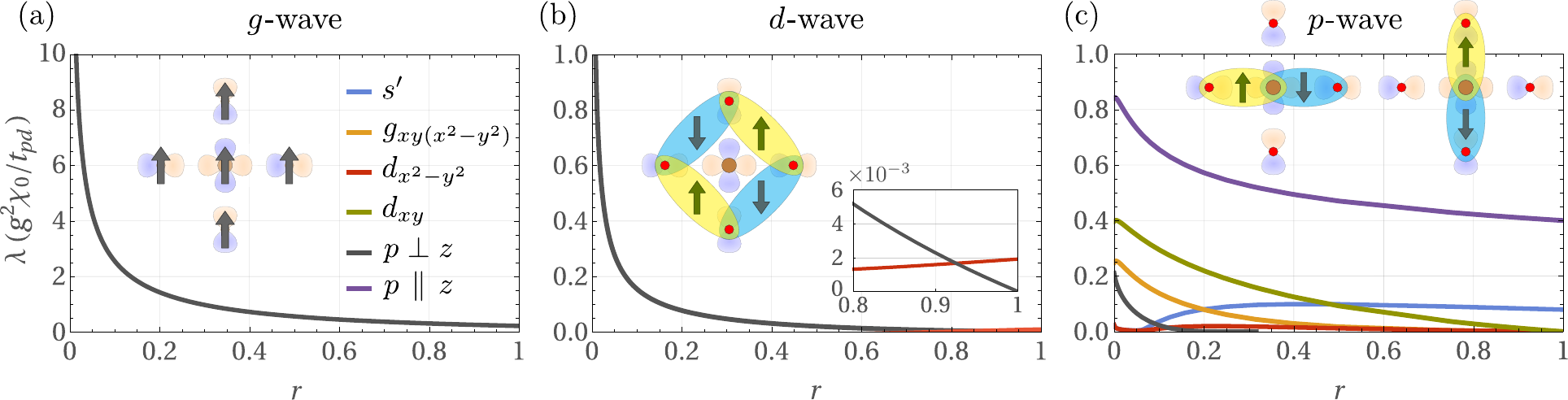}
\caption{Pairing eigenvalues $\lambda$ [Eq.~\eqref{eq:gapeq}] from spin fluctuations as a function of $r$ at fixed $\mu = 0.90 t_{pd}$.
In the presence of spin-orbit coupling, these secondary spin fluctuations are triggered by the orbital LC order parameters $\Phi_g$~(a), $\Phi_d$~(b), and $\vb{\Phi}_{p}$~(c).
By symmetry, the spins must fluctuate along the $z$ axis; the insets show on which sites or bonds they reside.
The even-parity states have a divergent pairing strength at the QCP, signalling strong triplet pairing.
In contrast, the odd-parity state is not significantly enhanced near the QCP.
The eigenvalues for spin-fluctuation-induced pairing are much larger than those promoted by the corresponding LC fluctuations in Fig.~\ref{fig:orbital-all}.
These spin eigenvalues will be reduced by the smallness of the spin-orbit coupling.
For the triplet states $p \perp z$ and $p \parallel z$, the Balian-Werthamer vector $\Delta_a$ is oriented along the $xy$-plane and the $z$-axis, respectively.}
\label{fig:spin-all}
\end{figure*}

Finally, $\vb{\Phi}_{p} = (\Phi_{p_{x}},\Phi_{p_{y}})$ is a two-component order parameter that describes a $p$-wave loop current, giving rise to a toroidal magnetic dipole moment.
It transforms under the $E_{u}$ irrep and thus has odd parity, $p_{\Phi}=-1$.
Its statistical mechanics is governed by a four-state clock model, a result that follows from a Landau expansion that includes quartic terms.
The four states for $\alpha = 0$, defined in Fig.~\ref{fig:Eu-orbital-alpha}, are shown in Fig.~\ref{fig:CuO2_plane}(d).
$\vb{\Phi}_{p}$ has a magneto-electric response, that is to say it can be polarized by crossed electric and magnetic fields according to $\mathcal{H}_{c} = - \kappa (\Phi_{p_x} B_{x} + \Phi_{p_y} B_{y}) E_{z}$.
A similar effect can be achieved by applying, instead of electric fields, time-varying currents along the $z$ axis.
As shown in Fig.~\ref{fig:orbital-all}(c), away from the QCP, we find that $\vb{\Phi}_{p}$ fluctuations result in weak extended $s$-wave superconductivity that is dominated by an angle-dependent gap function of the form $\Delta(\varphi) = \Delta_{0} + \Delta_{1} \cos(4\varphi)$ with $\abs{\Delta_{1}} \gg \abs{\Delta_{0}}$, yielding eight vertical line nodes.
In addition, there is a sub-leading weak $d_{x^{2}-y^{2}}$ pairing state which could only become dominant if one could approach smaller hole doping values without increasing the LC correlation length; see Fig.~\ref{fig:orbital-all}(f).
Most importantly, and in complete agreement with the general discussion for odd-parity LC states after Eq.~\eqref{eq:nearQCP}, the pairing eigenvalues turn strongly repulsive in all symmetry channels as one approaches the QCP, as signaled by the absence of any positive eigenvalue in Fig.~\ref{fig:orbital-all}(c) as $r \to 0$.
While the results in Fig.~\ref{fig:orbital-all}(c),(f) refer to $\alpha=0$, in Fig.~\ref{fig:Eu-orbital-alpha}(a) we show the impact of the parameter $\alpha$ on pairing. Recall that $\alpha$ parametrizes the relative weights between different paths connecting opposite \ce{O} orbitals; see Fig.~\ref{fig:Eu-orbital-alpha}(b)-(c).
The impact is clearly minor, consisting of the emergence of other weak subleading states for a range of $\alpha$ values and of the suppression of the leading state near $\alpha=\pi/4$.

Our analysis so far has considered only pure orbital magnetism.
Of course, in any system with spin-orbit coupling, spin fluctuations with the same symmetry as the LC patterns are expected to emerge~\cite{Klug2018,Christensen2022}.
Their pairing we analyze with a generalization of Eq.~\eqref{eq:gapeq} to spin exchange.
The degeneracy between the in-plane and out-of-plane triplet channels is now lifted by the non-trivial spin structure.
The results are shown in Figure~\ref{fig:spin-all}.
For even-parity LC order, we find strong pairing in triplet channels that will eventually dominate as $r \to 0$ over the weak singlet instabilities discussed earlier.
Conversely, for odd-parity LCs, spin fluctuations promote parametrically weak triplet pairing.
Hence the strongly repulsive behavior of the pairing interaction in the orbital sector cannot be offset by the contribution from spin fluctuations.
These results can, in fact, be derived from the relation Eq.~\eqref{eq:nearQCP}, adapted to spin-mediated pairing.
The crucial difference is that the conditions for the two behaviors in Eq.~\eqref{eq:nearQCP} are interchanged: now, forward scattering vanishes for odd parity ($p_{\Phi}=-1$).

\section*{Discussion}
In our analysis, we did not derive the existence of orbital magnetism or loop currents.
Instead, we started with the assumption that they exist and then explored the strength and type of superconductivity that they promote.
We find that away from the associated QCP, intra-unit-cell LC fluctuations can give rise to unconventional pairing.
Indeed, due to the negative-definiteness of the singlet Cooper channel interaction, $V_{+}(\vb{p},\vb{k}) < 0$, any $s$-wave solution that has no nodes, $\Delta_{+}(\vb{k}) > 0$, necessarily has negative eigenvalues $\lambda$ in Eq.~\eqref{eq:gapeq}.
Hence attractive pairing channels must be unconventional, if they exist.
However, in distinction to pairing mediated by spin-magnetic~\cite{Millis2001,Chubukov2003}, nematic~\cite{Lederer2015,Lederer2017,Klein2018}, or ferroelectric~\cite{Kozii2015,Klein2023} fluctuations, we do not find an enhancement of the pairing that is related to the vicinity of the QCP.
For even-parity loop currents, the weak orbital pairing behavior persists near the QCP and, in the presence of spin-orbit coupling, is overwhelmed by the symmetry-equivalent spin fluctuation that favor triplet pairing.
In contrast, odd-parity loop currents near the QCP are strongly pair-breaking in all symmetry channels, a behavior that cannot be circumvented even when one allows for spin-orbit coupling and spin fluctuations.
This behavior is consistent with the fact that in the ordered phase, a space-inversion and time-reversal symmetry breaking state would suppress the Cooper instability.
Hence, critical intra-unit-cell odd-parity LCs behave analogously to photons, where the coupling of a fermion current to the vector potential -- odd under parity and time reversal -- yields no superconductivity either.
Note that the absence of a strong attractive pairing interaction at the QCP justifies \textit{a posteriori} the weak-coupling analysis employed in this paper. 

With regards to the cuprate superconductors, we have studied them in the far-overdoped regime where complications relating to Mott physics, the pseudogap, and competing orders can be neglected~\cite{Kramer2019,Lee-Hone2020}.
We find that the odd-parity LC state $\vb{\Phi}_{p}$, which is the one most widely discussed in the cuprates, will at best give rise to extended $s$-wave pairing away from the QCP.
As one approaches the QCP, $\vb{\Phi}_{p}$ will not only fail to induce pairing in any channel, but will in fact suppress pairing that might arise from other collective modes.

In Ref.~\cite{Aji2010}, it was argued that pairing is caused by fluctuations of the conjugate momentum of $\vb{\Phi}_{p}$ and that this conjugate momentum has an essentially momentum-independent correlation function.
The conjugate momentum of $\vb{\Phi}_{p}$ must be odd under time reversal, even under parity, and transform like a magnetic dipole.
Hence, it transforms like $\Phi_{g}$ and is governed by the same form factor.
The analysis at the QCP is then formally identical to the case $r \sim 1$, where, as demonstrated in Fig.~\ref{fig:orbital-all}(a), we find weak $d_{xy}$-wave pairing.
We do not find the $d_{x^{2}-y^{2}}$ pairing state that was reported for fluctuations from $\Phi_{g}$~\cite{Aji2010,Varma2012}.
The reason is that the matrix element $f(\vb{p},\vb{k}) \propto \left[(\vb{k} \vcross \vb{p}) \vdot \vu{z}\right]^{2}$ used in Refs.~\cite{Aji2010,Varma2012} was estimated in the continuum limit, which ignores the fact that $f(\vb{p},\vb{k})$ vanishes when either $\vb{p}$ or $\vb{k}$ are at the high-symmetry Van Hove points $(\pi,0)$ or $(0,\pi)$; see Appendix~\ref{sec:bilinears}.
$\Phi_g$ therefore cannot take advantage of the high density of states near these Van Hove momenta.
Only for chemical potentials that yield very small electron pockets do we retrieve the degeneracy between $d_{xy}$ and $d_{x^{2}-y^{2}}$ pairing that follows from the continuum limit analysis~\cite{Aji2010,Varma2012}; see Fig.~\ref{fig:orbital-all}(d).
The continuum limit is, however, fully consistent with Eq.~\eqref{eq:nearQCP} for $p_{\Phi}=+1$, i.e., it would only give rise to weak pairing near the QCP.
While pairing due to the conjugate momentum is allowed and interesting, the order parameter itself should always couple directly to electrons and its much stronger pair-breaking tendency cannot, in our view, be ignored.

In conclusion, superconductivity due to this highly interesting state of matter is unlikely in general and in the cuprates in particular if we restrict ourselves to intra-unit-cell (i.e., $\vb{q} = \vb{0}$) ordering.
It is an interesting question of whether this remains the case if loop currents break additional translation symmetries, which has been proposed to take place in cuprates~\cite{Morr2001, Chakravarty2001, Varma2015, Varma2019, Bounoua2022, Bounoua2023} and in kagome superconductors~\cite{Mielke2022}.
For such staggered loop currents, the Cooper channel interaction is again uniformly repulsive, $V_{+}(\vb{p}, \vb{k}) < 0$, but now with a peak at a finite momentum transfer $\vb{Q}$, a behavior known to give rise to unconventional superconductivity~\cite{scalapino1995case,Maiti2013}.
We find that such pairing due to staggered LCs strengthens as the QCP is approached as there are no generic symmetries, like parity or time-reversal, that suppress it.
For the electronic structure of the cuprates, we obtain that $d$-wave and $p$-wave LCs with $\vb{Q} = (\pi,\pi)$ both favor strong $d_{x^2-y^2}$-wave pairing, whereas $g$-wave LCs do not because of the suppression of $f(\vb{p},\vb{k})$ at the van Hove points.

\emph{Acknowledgments:}
We are grateful to A.\ V.\ Chubukov, T.\ P.\ Devereaux, S.\ A.\ Kivelson, and C.\ M.\ Varma for useful discussions.
Work at KIT was supported by the Deutsche Forschungsgemeinschaft (DFG, German Research Foundation) - TRR 288-422213477 Elasto-Q-Mat (project A07). R.M.F.\ was supported by the Air Force Office of Scientific Research under Award No.\ FA9550-21-1-0423. R.M.F.\ also acknowledges a Mercator Fellowship from the German Research Foundation (DFG) through TRR 288, 422213477 Elasto-Q-Mat. J.S.\ and R.M.F.\ acknowledge the hospitality of KITP, where part of this work was done. KITP is supported in part by the National Science Foundation under Grant No.\ NSF PHY-1748958 and NSF PHY-2309135.

\appendix

\section{Loop-current exchange} \label{sec:LC-exchange}
\begin{figure}[t]
\centering
\includegraphics[width=\linewidth]{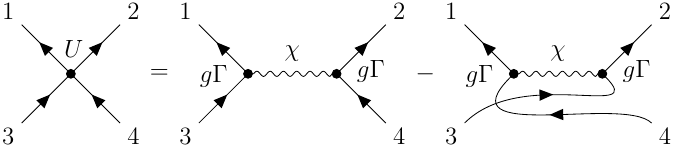}
\caption{Diagram of the antisymmetrized four-fermion interaction that is mediated by a bosonic collective mode. The bosonic propagator is denoted with a wavy line.}
\label{fig:diagram}
\end{figure}

The fully antisymmetrized interaction due to exchange of LC fluctuations that couple to the fermionic bilinear of Eq.~\eqref{eq:phi} equals $U_{1234} = - (V_{1234} - V_{1243})$, where $i \equiv (\vb{k}_i, \alpha_i)$, $\alpha_{i}$ is the combined orbital and spin index, and
\begin{equation}
V_{1234} = g^{2} \chi(\vb{k}_{1}-\vb{k}_{3}) \left[\Gamma_{\vb{k}_{1},\vb{k}_{3}}\right]_{\alpha_{1}\alpha_{3}} \left[\Gamma_{\vb{k}_{2}, \vb{k}_{4}}\right]_{\alpha_{2}\alpha_{4}}.
\end{equation}
See Fig.~\ref{fig:diagram} for the diagram.
The pairing instability is determined by the Cooper channel interaction $\tilde{U}^{\alpha_1\alpha_2}_{\alpha_3\alpha_4}(\vb{k}_1,\vb{k}_3) \equiv U_{\alpha_1\alpha_2\alpha_3\alpha_4}(\vb{k}_1,-\vb{k}_1,\vb{k}_3,-\vb{k}_3)$ through the linearized gap equation~\cite{Leggett2006, Sigrist2005}:
\begin{align}
[\Delta_{\vb{p}}]_{\alpha_1\alpha_2} = - \frac{1}{2N} \sum_{\vb{k}\alpha_3\alpha_4} &\tilde{U}^{\alpha_1\alpha_2}_{\alpha_3\alpha_4}(\vb{p},\vb{k}) \sum_{n} \frac{\tanh\tfrac{1}{2}\beta\varepsilon_{\vb{k}n}}{2\varepsilon_{\vb{k}n}} \notag \\
&\quad \times \left[P_{\vb{k}n} \Delta_{\vb{k}} P_{-\vb{k}n}^{\intercal}\right]_{\alpha_3\alpha_4}.
\end{align}
Here $n$ is the bands index, $\varepsilon_{\vb{k}n}$ the band dispersion, and $P_{\vb{k}n} = \sum_s \left\lvert u_{\vb{k}ns}\right\rangle \left\langle u_{\vb{k}ns}\right\rvert$ projects onto the Bloch states of the band.
The gap matrix we express in terms of a Balian-Werthamer vector $\Delta_{a}(\vb{p})$:
\begin{equation}
[\Delta_{\vb{p}}]_{\alpha_1\alpha_2} = \sum_{a=0}^{3} \Delta_{a}(\vb{p})\left[P_{\vb{p}}^{a} \, \iu \Pauli^y\right]_{\alpha_1\alpha_2},
\end{equation}
where $P_{\vb{p}}^{a} = \sum_{ss'} u_{\vb{p}ns} \Pauli^a_{ss'} u_{\vb{p}ns'}^{\dag}$ for that $n$ whose $\varepsilon_{\vb{p}n} = 0$.
Projecting onto a single Fermi surface for purely orbital $\Gamma_{\vb{p},\vb{k}} = \gamma_{\vb{p},\vb{k}} \otimes \Pauli^{0}$ under the assumption of no spin-orbit coupling yields the linearized gap equation~\eqref{eq:gapeq}.

The matrix element $f(\vb{p},\vb{k}) = \abs{u_{\vb{p}}^{\dag} \gamma_{\vb{p},\vb{k}} u_{\vb{k}}}^2$ of Eq.~\eqref{eq:aux_pairing_int} may vanish for $\vb{p} = \pm \vb{k}$, depending on the parity and time-reversal sign of the order parameter~\eqref{eq:phi}.
Under inversion symmetry $\gamma_{\vb{p},\vb{k}} \overset{P}{\to} p_{\Phi} \gamma_{-\vb{p},-\vb{k}}$, where $p_{\Phi}$ is the parity of $\Phi$. Since LCs are odd under time-reversal, $\gamma_{\vb{p},\vb{k}} \overset{\Theta}{\to} - \gamma_{-\vb{p},-\vb{k}}^{*}$. If we further use the transformation properties of orbital Bloch functions $u_{\vb{k}} \overset{P}{\to} u_{-\vb{k}} = u_{\vb{k}}^{*}$ under these same symmetries, we find:
\begin{equation}
u_{\vb{k}}^{\dag} \gamma_{\vb{k},\vb{p}} u_{\vb{p}} = - p_{\Phi} u_{\vb{p}}^{\dag} \gamma_{\vb{p},\vb{k}} u_{\vb{k}}.
\end{equation}
Hence for even-parity ($p_{\Phi} = +1$) LCs, the matrix element $f(\vb{p},\vb{k})$ vanishes at $\vb{p} = \vb{k}$, yielding Eq.~\eqref{eq:nearQCP}.
Due to time-reversal oddness of LCs, $u_{-\vb{k}}^{\dag} \gamma_{-\vb{k},-\vb{p}} u_{-\vb{p}} = - u_{\vb{p}}^{\dag} \gamma_{\vb{p},\vb{k}} u_{\vb{k}}$ so $f(-\vb{k},\vb{k}) = 0$ also vanishes.

\section{Electronic structure of cuprates} \label{sec:CuO2-TBA-model}
In cuprates, the states closest to the Fermi level primarily derive from anti-bonding hybridization between \ce{Cu}:$3d_{x^2-y^2}$ orbitals and \ce{O}:$2p_{x,y}$ orbitals oriented along the ligands~\cite{Pickett1989, Dagotto1994, Pellegrin1993, Varma1987}; see Fig.~\ref{fig:CuO2_plane}(a).
These orbitals are the basis of the three-band tight-binding model~\cite{Varma1987, Emery1987, Emery1988, Littlewood1989, Gaididei1988, Scalettar1991} that we employ in our calculation.
In the orbital basis $c_{\vb{k}\sigma} = \left(d_{\vb{k}\sigma}, p_{x,\vb{k}\sigma}, p_{y,\vb{k}\sigma}\right)^{\intercal}$, the three-band Hamiltonian takes the form
\begin{align}
H_{\vb{k}} &= \begin{pmatrix}
h_d(\vb{k}) & h_{pd}(\vb{k}) & - h_{pd}(\vb{\tilde{k}}) \\
& h_p(\vb{k}) & h_{pp}(\vb{k}) \\
\mathrm{c.c.} & & h_p(\vb{\tilde{k}})
\end{pmatrix}, \label{eq:3band-Haml}
\end{align}
where $\vb{k} = (k_x,k_y)$, $\vb{\tilde{k}} = (k_y,k_x)$, and
\begin{align}
\begin{aligned}
h_d(\vb{k}) &= \epsilon_d - \mu, \\
h_p(\vb{k}) &= \epsilon_p + 2 t_{pp}' \cos k_x - \mu, \\
h_{pd}(\vb{k}) &= t_{pd} (1 - \Elr^{- \iu k_x}), \\
h_{pp}(\vb{k}) &= - t_{pp} (1 - \Elr^{\iu k_x}) (1 - \Elr^{- \iu k_y}).
\end{aligned}
\end{align}
Typical values for the tight-binding parameters used in the literature are~\cite{Photopoulos2019} $(\epsilon_d - \epsilon_p) / t_{pd} \in [2.5, 3.5]$, $t_{pp} / t_{pd} \in [0.5, 0.6]$, and $t_{pp}' / t_{pd} \approx 0$ with $t_{pd} \in [1.2, 1.5]~\si{\electronvolt}$.
$t_{pp}'$ is not really negligible~\cite{Andersen1995, Pavarini2001, Kent2008}, although it is often assumed to be.
The importance of $t_{pp}'$ for LCs was emphasized in Ref.~\cite{Weber2014}.
We have considered eight different parameter sets that cover a wide range of physically reasonable possibilities~\cite{Pickett1989, Pavarini2001, Kent2008, Weber2014, Photopoulos2019} and that reproduce the ARPES Fermi surface shapes~\cite{Plate2005, Yoshida2006, Peets2007, Horio2018}.
Our results have turned out to be insensitive to these changes in the one-particle Hamiltonian.
All results shown or quoted in the paper are for the representative parameter set $\epsilon_d - \epsilon_p = 3 t_{pd}$, $t_{pp} = 0.6 t_{pd}$, $t'_{pp} = 0.5 t_{pd}$, and $\mu = 0.9 t_{pd}$ with $\epsilon_d = 0$.

\section{Classification of fermionic bilinears} \label{sec:bilinears}
Because of the non-trivial Wyckoff positions of the oxygen atoms, some point group operations (e.g., $90^{\circ}$ rotations and parity) map orbitals between different primitive unit cells.
In momentum space, the corresponding unitary matrices therefore acquire $\vb{k}$-dependent phases.
For classification purposes, it is more convenient if the orbital and momentum dependencies of the point group matrices do not mix.
Instead of the three-component spinor $c_{\sigma}(\vb{R}) = \left(d_{\sigma}(\vb{R}), p_{x,\sigma}(\vb{R}+\tfrac{1}{2}\vu{x}), p_{y,\sigma}(\vb{R}+\frac{1}{2}\vu{y})\right)^{\intercal}$, we therefore employ an extended five-component spinor $a_{\sigma}(\vb{R}) = \big(d_{\sigma}(\vb{R}), p_{x,\sigma}(\vb{R}+\tfrac{1}{2}\vu{x}), p_{y,\sigma}(\vb{R}+\frac{1}{2}\vu{y}), p_{x,\sigma}(\vb{R}-\tfrac{1}{2}\vu{x}), p_{y,\sigma}(\vb{R}-\frac{1}{2}\vu{y})\big)^{\intercal}$ that is related to the primitive spinor through $a_{\vb{k}\sigma} = W_{\vb{k}} c_{\vb{k}\sigma}$, where
\begin{align}
W_{\vb{k}} &= \begin{pmatrix}
1 & 0 & 0 \\
0 & 1 & 0 \\
0 & 0 & 1 \\
0 & \Elr^{- \iu k_x} & 0 \\
0 & 0 & \Elr^{- \iu k_y}
\end{pmatrix}. \label{eq:W-proj-mat}
\end{align}
The corresponding extended unit cell is shaded blue in Fig.~\ref{fig:CuO2_plane}(a).

In the extended basis, point group transformation matrices no longer depend on $\vb{k}$.
Hence LCs are classified by imaginary Hermitian $5 \times 5$ matrices.
In total, there are $10$ linearly independent matrices which we classify into irreps of the $D_{4h}$ point group~\cite{Dresselhaus2007} below:
\begin{align*}
&\Lambda^{A_{1g}} = \frac{1}{2} \begin{pmatrix}
0 & \!\!-\iu & \iu & \iu & \!\!-\iu \\
\iu & 0 & 0 & 0 & 0 \\
\!\!-\iu & 0 & 0 & 0 & 0 \\
\!\!-\iu & 0 & 0 & 0 & 0 \\
\iu & 0 & 0 & 0 & 0 \\
\end{pmatrix},
\Lambda^{A_{2g}} = \frac{1}{2} \begin{pmatrix}
0 & 0 & 0 & 0 & 0 \\
0 & 0 & \!\!-\iu & 0 & \!\!-\iu \\
0 & \iu & 0 & \iu & 0 \\
0 & 0 & \!\!-\iu & 0 & \!\!-\iu \\
0 & \iu & 0 & \iu & 0 \\
\end{pmatrix}, \\
&\Lambda^{B_{1g}}_1 = \frac{1}{2} \begin{pmatrix}
0 & \!\!-\iu & \!\!-\iu & \iu & \iu \\
\iu & 0 & 0 & 0 & 0 \\
\iu & 0 & 0 & 0 & 0 \\
\!\!-\iu & 0 & 0 & 0 & 0 \\
\!\!-\iu & 0 & 0 & 0 & 0 \\
\end{pmatrix},
\Lambda^{B_{1g}}_2 = \frac{1}{2} \begin{pmatrix}
0 & 0 & 0 & 0 & 0 \\
0 & 0 & \iu & 0 & \!\!-\iu \\
0 & \!\!-\iu & 0 & \iu & 0 \\
0 & 0 & \!\!-\iu & 0 & \iu \\
0 & \iu & 0 & \!\!-\iu & 0 \\
\end{pmatrix}, \\
&\Lambda^{E_{u}}_{1,x}\!=\!\frac{1}{\sqrt{2}}\! \begin{pmatrix}
0 & \!\!-\iu & 0 & \!\!-\iu & 0 \\
\iu & 0 & 0 & 0 & 0 \\
0 & 0 & 0 & 0 & 0 \\
\iu & 0 & 0 & 0 & 0 \\
0 & 0 & 0 & 0 & 0 \\
\end{pmatrix},
\Lambda^{E_{u}}_{1,y}\!=\!\frac{1}{\sqrt{2}}\! \begin{pmatrix}
0 & 0 & \iu & 0 & \iu \\
0 & 0 & 0 & 0 & 0 \\
\!\!-\iu & 0 & 0 & 0 & 0 \\
0 & 0 & 0 & 0 & 0 \\
\!\!-\iu & 0 & 0 & 0 & 0 \\
\end{pmatrix}, \\
&\Lambda^{E_{u}}_{2,x} = \mkern15mu \begin{pmatrix}
0 & 0 & 0 & 0 & 0 \\
0 & 0 & 0 & \!\!-\iu & 0 \\
0 & 0 & 0 & 0 & 0 \\
0 & \iu & 0 & 0 & 0 \\
0 & 0 & 0 & 0 & 0 \\
\end{pmatrix}, \mkern11mu
\Lambda^{E_{u}}_{2,y} = \mkern15mu \begin{pmatrix}
0 & 0 & 0 & 0 & 0 \\
0 & 0 & 0 & 0 & 0 \\
0 & 0 & 0 & 0 & \!\!-\iu \\
0 & 0 & 0 & 0 & 0 \\
0 & 0 & \iu & 0 & 0 \\
\end{pmatrix}, \\
&\Lambda^{E_{u}}_{3,x} = \frac{1}{2} \begin{pmatrix}
0 & 0 & 0 & 0 & 0 \\
0 & 0 & \iu & 0 & \!\!-\iu \\
0 & \!\!-\iu & 0 & \!\!-\iu & 0 \\
0 & 0 & \iu & 0 & \!\!-\iu \\
0 & \iu & 0 & \iu & 0 \\
\end{pmatrix},
\Lambda^{E_{u}}_{3,y} = \frac{1}{2} \begin{pmatrix}
0 & 0 & 0 & 0 & 0 \\
0 & 0 & \!\!-\iu & 0 & \!\!-\iu \\
0 & \iu & 0 & \!\!-\iu & 0 \\
0 & 0 & \iu & 0 & \iu \\
0 & \iu & 0 & \!\!-\iu & 0 \\
\end{pmatrix}.
\end{align*}
These matrices determine the form factors of the LC order parameter [Eq.~\eqref{eq:phi}] via $\Gamma_{\vb{k},\vb{p}} = W_{\vb{k}}^{\dag} \Lambda W_{\vb{p}} \otimes \Pauli^0$.
Hence one would expect a total of $10$ possible LC states.
In the next appendix, we show that charge conservation and the fact that spontaneous global currents are forbidden in purely electronic systems reduce the total number down to six.

The enhanced density of states near the Van Hove points $\vb{k}_{X} = (\pi,0)$ and $\vb{k}_{Y} = (0,\pi)$ can be important for pairing.
This turns out to not be the case for $\Phi_g$-mediated pairing associated with $A_{2g}$ LCs.
To see why, let us consider $\vb{k}_{X}$.
Parity implies that the band Hamiltonian is block-diagonal at this momentum, with the $(d, p_{x})$ block decoupled from $p_{y}$.
Consequently, the form factor $W_{\vb{p}}^{\dag} \Lambda^{A_{2g}} W_{\vb{k}_{X}}$ between $\vb{k}_{X}$ and an arbitrary momentum $\vb{p}$ vanishes after projection onto the conduction band.

\section{Bloch and Kirchhoff constraints} \label{sec:current-constraints}
\begin{figure}[t]
\centering
\includegraphics[width=\linewidth]{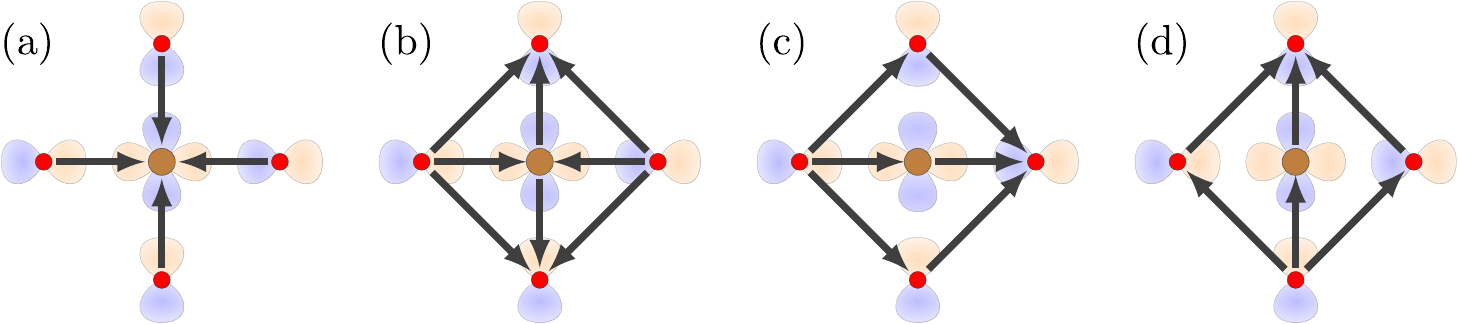}
\caption{(a) and (b): Local currents that would result in an accumulation of charge on $d$ and $p_y$ orbitals, respectively. (c),(d): Global currents that would violate Bloch's theorem.}
\label{fig:Bloch-Kirchhoff}
\end{figure}

A LC state is allowed only if its ordered state is consistent with the continuity equation (Kirchhoff's law) and with the Bloch constraint~\cite{Bohm1949,Ohashi1996,Watanabe2022} that no global current can flow; see Fig.~\ref{fig:Bloch-Kirchhoff}.
A microscopic theory that properly derives such a state would naturally obey these conditions.
In our phenomenological treatment, we must impose these constraints explicitly.

\emph{Bloch constraints:}
Using the extended basis, the global current operator can be written as
\begin{equation}
\vb{j} = \frac{1}{N} \sum_{\vb{R}ij} a_{i\sigma}^{\dag}(\vb{R}) \vb{J}_{ij} a_{j\sigma}(\vb{R}).
\end{equation}
Here $i,j$ are orbital indices and $(\vb{J})_{ij} = - \iu \left(\vb{r}_{i} - \vb{r}_{j}\right) t_{ij}$, where $\vb{r}_{i}$ are the basis vectors of the five extended unit cell atoms.
The current matrix $\vb{J}$ can be expressed in terms of time-reversal-odd $E_u$ matrices:
\begin{equation}
J_w = - \frac{1}{\sqrt{2}} t_{pd} \Lambda^{E_u}_{1,w} + t_{pp}' \Lambda^{E_u}_{2,w} + t_{pp} \Lambda^{E_u}_{3,w},
\end{equation}
where $w = x$ or $y$.
Hence a finite global current can only be induced by $p$-wave LC order that transforms under the same $E_u$ irrep.
Consider a small, but finite, order parameter $\vb{\Phi}_{p}$ that couples to fermions through the linear combination of form factors
\begin{equation}
\Gamma_{\vb{k},\vb{p}} = W_{\vb{k}}^{\dag} \left(c_1 \Lambda^{E_u}_{1,w} + c_2 \Lambda^{E_u}_{2,w} + c_3 \Lambda^{E_u}_{3,w}\right) W_{\vb{p}}. \label{eq:Gamma_Eu}
\end{equation}
Linear response theory then yields the constraint
\begin{equation}
\frac{\delta \langle j_{w}\rangle}{\delta \Phi_{p,w}} = - g \, \vb{h} \vdot \vb{c} = 0,
\end{equation}
where $\vb{h} = (h_1, h_2, h_3)$ are the linear response coefficients obtained by evaluating the current expectation value and $\vb{c} = (c_1, c_2, c_3)$ specify the bilinear~\eqref{eq:Gamma_Eu}.
$\vb{h} / \abs{\vb{h}}$ depends weakly on chemical potential and for $\mu = 0.9 t_{pd}$ equals $(0.85, 0.31, -0.43)$.
After enforcing the above constraint, a one-parameter family of $E_u$ bilinears $\vb{c} = \vu{h}_c \cos \alpha + \vu{h}_s \sin \alpha $ remains.
Here $\vb{h}_c = \vb{h} \vcross (0,1,0)$, $\vu{h}_c = \vb{h}_c / \abs{\vb{h}_c}$, $\vb{h}_s = \vb{h} \vcross \vb{h}_c$, and $\vu{h}_s = \vb{h}_s / \abs{\vb{h}_s}$.

\emph{Kirchhoff constraints:}
Local charge conservation entails that $\dot{n}_{i} = \sum_{j} G_{ij}$ where $n_{i}$ is the charge on site $i$ and $G_{ij} = - G_{ji}$ is the current flowing from site $j$ to $i$.
For a steady state, it must hold that $\langle \dot{n}_{i} \rangle = 0$.
One easily finds that $G^{A_{1g}} = 2 t_{pd} \Lambda^{A_{1g}}_6$ currents change the copper $n_d$ and total oxygen $n_{p_x}+n_{p_y}$ charge, whereas $G^{B_{1g}} = - 2 t_{pd} \Lambda^{B_{1g}}_4 + 4 t_{pp} \Lambda^{B_{1g}}_5$ currents change $n_{p_x}-n_{p_y}$.
Hence, currents associated with $G^{A_{1g}}$ and $G^{B_{1g}}$ must vanish.
This implies that there cannot be a LC of $A_{1g}$ symmetry and that the two coefficients $c_1$ and $c_2$ in the bilinear
\begin{align}
\Gamma_{\vb{k},\vb{p}} &= W_{\vb{k}}^{\dag} \left(c_1 \Lambda^{B_{1g}}_1 + c_2 \Lambda^{B_{1g}}_2\right) W_{\vb{p}}
\end{align}
are not independent.
For $\mu = 0.9t_{pd}$ we find that $c_1/c_2 = 0.72$; this ratio depends weakly on $\mu$.

\bibliography{cuprates-references}

\begin{thebibliography}{80}%
\makeatletter
\providecommand \@ifxundefined [1]{%
 \@ifx{#1\undefined}
}%
\providecommand \@ifnum [1]{%
 \ifnum #1\expandafter \@firstoftwo
 \else \expandafter \@secondoftwo
 \fi
}%
\providecommand \@ifx [1]{%
 \ifx #1\expandafter \@firstoftwo
 \else \expandafter \@secondoftwo
 \fi
}%
\providecommand \natexlab [1]{#1}%
\providecommand \enquote  [1]{``#1''}%
\providecommand \bibnamefont  [1]{#1}%
\providecommand \bibfnamefont [1]{#1}%
\providecommand \citenamefont [1]{#1}%
\providecommand \href@noop [0]{\@secondoftwo}%
\providecommand \href [0]{\begingroup \@sanitize@url \@href}%
\providecommand \@href[1]{\@@startlink{#1}\@@href}%
\providecommand \@@href[1]{\endgroup#1\@@endlink}%
\providecommand \@sanitize@url [0]{\catcode `\\12\catcode `\$12\catcode
  `\&12\catcode `\#12\catcode `\^12\catcode `\_12\catcode `\%12\relax}%
\providecommand \@@startlink[1]{}%
\providecommand \@@endlink[0]{}%
\providecommand \url  [0]{\begingroup\@sanitize@url \@url }%
\providecommand \@url [1]{\endgroup\@href {#1}{\urlprefix }}%
\providecommand \urlprefix  [0]{URL }%
\providecommand \Eprint [0]{\href }%
\providecommand \doibase [0]{https://doi.org/}%
\providecommand \selectlanguage [0]{\@gobble}%
\providecommand \bibinfo  [0]{\@secondoftwo}%
\providecommand \bibfield  [0]{\@secondoftwo}%
\providecommand \translation [1]{[#1]}%
\providecommand \BibitemOpen [0]{}%
\providecommand \bibitemStop [0]{}%
\providecommand \bibitemNoStop [0]{.\EOS\space}%
\providecommand \EOS [0]{\spacefactor3000\relax}%
\providecommand \BibitemShut  [1]{\csname bibitem#1\endcsname}%
\let\auto@bib@innerbib\@empty
\bibitem [{\citenamefont {Bohm}(1949)}]{Bohm1949}%
  \BibitemOpen
  \bibfield  {author} {\bibinfo {author} {\bibfnamefont {D.}~\bibnamefont
  {Bohm}},\ }\bibfield  {title} {\bibinfo {title} {Note on a theorem of {B}loch
  concerning possible causes of superconductivity},\ }\href
  {https://doi.org/10.1103/PhysRev.75.502} {\bibfield  {journal} {\bibinfo
  {journal} {Phys. Rev.}\ }\textbf {\bibinfo {volume} {75}},\ \bibinfo {pages}
  {502} (\bibinfo {year} {1949})}\BibitemShut {NoStop}%
\bibitem [{\citenamefont {Ohashi}\ and\ \citenamefont
  {Momoi}(1996)}]{Ohashi1996}%
  \BibitemOpen
  \bibfield  {author} {\bibinfo {author} {\bibfnamefont {Y.}~\bibnamefont
  {Ohashi}}\ and\ \bibinfo {author} {\bibfnamefont {T.}~\bibnamefont {Momoi}},\
  }\bibfield  {title} {\bibinfo {title} {On the {B}loch theorem concerning
  spontaneous electric current},\ }\href {https://doi.org/10.1143/JPSJ.65.3254}
  {\bibfield  {journal} {\bibinfo  {journal} {J. Phys. Soc. Japan}\ }\textbf
  {\bibinfo {volume} {65}},\ \bibinfo {pages} {3254} (\bibinfo {year}
  {1996})}\BibitemShut {NoStop}%
\bibitem [{\citenamefont {Watanabe}(2022)}]{Watanabe2022}%
  \BibitemOpen
  \bibfield  {author} {\bibinfo {author} {\bibfnamefont {H.}~\bibnamefont
  {Watanabe}},\ }\bibfield  {title} {\bibinfo {title} {{B}loch theorem in the
  presence of an additional conserved charge},\ }\href
  {https://doi.org/10.1103/PhysRevResearch.4.013043} {\bibfield  {journal}
  {\bibinfo  {journal} {Phys. Rev. Res.}\ }\textbf {\bibinfo {volume} {4}},\
  \bibinfo {pages} {013043} (\bibinfo {year} {2022})}\BibitemShut {NoStop}%
\bibitem [{\citenamefont {Varma}(1997)}]{Varma1997}%
  \BibitemOpen
  \bibfield  {author} {\bibinfo {author} {\bibfnamefont {C.~M.}\ \bibnamefont
  {Varma}},\ }\bibfield  {title} {\bibinfo {title} {Non-{F}ermi-liquid states
  and pairing instability of a general model of copper oxide metals},\ }\href
  {https://doi.org/10.1103/PhysRevB.55.14554} {\bibfield  {journal} {\bibinfo
  {journal} {Phys. Rev. B}\ }\textbf {\bibinfo {volume} {55}},\ \bibinfo
  {pages} {14554} (\bibinfo {year} {1997})}\BibitemShut {NoStop}%
\bibitem [{\citenamefont {Simon}\ and\ \citenamefont
  {Varma}(2003)}]{Simon2003}%
  \BibitemOpen
  \bibfield  {author} {\bibinfo {author} {\bibfnamefont {M.~E.}\ \bibnamefont
  {Simon}}\ and\ \bibinfo {author} {\bibfnamefont {C.~M.}\ \bibnamefont
  {Varma}},\ }\bibfield  {title} {\bibinfo {title} {Symmetry considerations for
  the detection of second-harmonic generation in cuprates in the pseudogap
  phase},\ }\href {https://doi.org/10.1103/PhysRevB.67.054511} {\bibfield
  {journal} {\bibinfo  {journal} {Phys. Rev. B}\ }\textbf {\bibinfo {volume}
  {67}},\ \bibinfo {pages} {054511} (\bibinfo {year} {2003})}\BibitemShut
  {NoStop}%
\bibitem [{\citenamefont {Varma}(2006)}]{Varma2006}%
  \BibitemOpen
  \bibfield  {author} {\bibinfo {author} {\bibfnamefont {C.~M.}\ \bibnamefont
  {Varma}},\ }\bibfield  {title} {\bibinfo {title} {Theory of the pseudogap
  state of the cuprates},\ }\href {https://doi.org/10.1103/PhysRevB.73.155113}
  {\bibfield  {journal} {\bibinfo  {journal} {Phys. Rev. B}\ }\textbf {\bibinfo
  {volume} {73}},\ \bibinfo {pages} {155113} (\bibinfo {year}
  {2006})}\BibitemShut {NoStop}%
\bibitem [{\citenamefont {Aji}\ \emph {et~al.}(2010)\citenamefont {Aji},
  \citenamefont {Shekhter},\ and\ \citenamefont {Varma}}]{Aji2010}%
  \BibitemOpen
  \bibfield  {author} {\bibinfo {author} {\bibfnamefont {V.}~\bibnamefont
  {Aji}}, \bibinfo {author} {\bibfnamefont {A.}~\bibnamefont {Shekhter}},\ and\
  \bibinfo {author} {\bibfnamefont {C.~M.}\ \bibnamefont {Varma}},\ }\bibfield
  {title} {\bibinfo {title} {Theory of the coupling of quantum-critical
  fluctuations to fermions and ${d}$-wave superconductivity in cuprates},\
  }\href {https://doi.org/10.1103/PhysRevB.81.064515} {\bibfield  {journal}
  {\bibinfo  {journal} {Phys. Rev. B}\ }\textbf {\bibinfo {volume} {81}},\
  \bibinfo {pages} {064515} (\bibinfo {year} {2010})}\BibitemShut {NoStop}%
\bibitem [{\citenamefont {Aji}\ and\ \citenamefont {Varma}(2007)}]{Aji2007}%
  \BibitemOpen
  \bibfield  {author} {\bibinfo {author} {\bibfnamefont {V.}~\bibnamefont
  {Aji}}\ and\ \bibinfo {author} {\bibfnamefont {C.~M.}\ \bibnamefont
  {Varma}},\ }\bibfield  {title} {\bibinfo {title} {Spin order accompanying
  loop-current order in cuprate superconductors},\ }\href
  {https://doi.org/10.1103/PhysRevB.75.224511} {\bibfield  {journal} {\bibinfo
  {journal} {Phys. Rev. B}\ }\textbf {\bibinfo {volume} {75}},\ \bibinfo
  {pages} {224511} (\bibinfo {year} {2007})}\BibitemShut {NoStop}%
\bibitem [{\citenamefont {Varma}\ \emph {et~al.}(1987)\citenamefont {Varma},
  \citenamefont {Schmitt-Rink},\ and\ \citenamefont {Abrahams}}]{Varma1987}%
  \BibitemOpen
  \bibfield  {author} {\bibinfo {author} {\bibfnamefont {C.}~\bibnamefont
  {Varma}}, \bibinfo {author} {\bibfnamefont {S.}~\bibnamefont
  {Schmitt-Rink}},\ and\ \bibinfo {author} {\bibfnamefont {E.}~\bibnamefont
  {Abrahams}},\ }\bibfield  {title} {\bibinfo {title} {Charge transfer
  excitations and superconductivity in “ionic” metals},\ }\href
  {https://doi.org/https://doi.org/10.1016/0038-1098(87)90407-8} {\bibfield
  {journal} {\bibinfo  {journal} {Solid State Commun.}\ }\textbf {\bibinfo
  {volume} {62}},\ \bibinfo {pages} {681} (\bibinfo {year} {1987})}\BibitemShut
  {NoStop}%
\bibitem [{\citenamefont {Fauqu\'{e}}\ \emph {et~al.}(2006)\citenamefont
  {Fauqu\'{e}}, \citenamefont {Sidis}, \citenamefont {Hinkov}, \citenamefont
  {Pailh\`{e}s}, \citenamefont {Lin}, \citenamefont {Chaud},\ and\
  \citenamefont {Bourges}}]{Fauque2006}%
  \BibitemOpen
  \bibfield  {author} {\bibinfo {author} {\bibfnamefont {B.}~\bibnamefont
  {Fauqu\'{e}}}, \bibinfo {author} {\bibfnamefont {Y.}~\bibnamefont {Sidis}},
  \bibinfo {author} {\bibfnamefont {V.}~\bibnamefont {Hinkov}}, \bibinfo
  {author} {\bibfnamefont {S.}~\bibnamefont {Pailh\`{e}s}}, \bibinfo {author}
  {\bibfnamefont {C.~T.}\ \bibnamefont {Lin}}, \bibinfo {author} {\bibfnamefont
  {X.}~\bibnamefont {Chaud}},\ and\ \bibinfo {author} {\bibfnamefont
  {P.}~\bibnamefont {Bourges}},\ }\bibfield  {title} {\bibinfo {title}
  {Magnetic order in the pseudogap phase of high-${T}_{c}$ superconductors},\
  }\href {https://doi.org/10.1103/PhysRevLett.96.197001} {\bibfield  {journal}
  {\bibinfo  {journal} {Phys. Rev. Lett.}\ }\textbf {\bibinfo {volume} {96}},\
  \bibinfo {pages} {197001} (\bibinfo {year} {2006})}\BibitemShut {NoStop}%
\bibitem [{\citenamefont {Mook}\ \emph {et~al.}(2008)\citenamefont {Mook},
  \citenamefont {Sidis}, \citenamefont {Fauqu\'e}, \citenamefont {Bal\'edent},\
  and\ \citenamefont {Bourges}}]{Bourges2008}%
  \BibitemOpen
  \bibfield  {author} {\bibinfo {author} {\bibfnamefont {H.~A.}\ \bibnamefont
  {Mook}}, \bibinfo {author} {\bibfnamefont {Y.}~\bibnamefont {Sidis}},
  \bibinfo {author} {\bibfnamefont {B.}~\bibnamefont {Fauqu\'e}}, \bibinfo
  {author} {\bibfnamefont {V.}~\bibnamefont {Bal\'edent}},\ and\ \bibinfo
  {author} {\bibfnamefont {P.}~\bibnamefont {Bourges}},\ }\bibfield  {title}
  {\bibinfo {title} {Observation of magnetic order in a superconducting
  \ce{YBa2Cu3O_{6.6}} single crystal using polarized neutron scattering},\
  }\href {https://doi.org/10.1103/PhysRevB.78.020506} {\bibfield  {journal}
  {\bibinfo  {journal} {Phys. Rev. B}\ }\textbf {\bibinfo {volume} {78}},\
  \bibinfo {pages} {020506} (\bibinfo {year} {2008})}\BibitemShut {NoStop}%
\bibitem [{\citenamefont {Li}\ \emph {et~al.}(2008)\citenamefont {Li},
  \citenamefont {Bal{\'e}dent}, \citenamefont {Bari{\v{s}}i{\'{c}}},
  \citenamefont {Cho}, \citenamefont {Fauqu{\'e}}, \citenamefont {Sidis},
  \citenamefont {Yu}, \citenamefont {Zhao}, \citenamefont {Bourges},\ and\
  \citenamefont {Greven}}]{Greven2008}%
  \BibitemOpen
  \bibfield  {author} {\bibinfo {author} {\bibfnamefont {Y.}~\bibnamefont
  {Li}}, \bibinfo {author} {\bibfnamefont {V.}~\bibnamefont {Bal{\'e}dent}},
  \bibinfo {author} {\bibfnamefont {N.}~\bibnamefont {Bari{\v{s}}i{\'{c}}}},
  \bibinfo {author} {\bibfnamefont {Y.}~\bibnamefont {Cho}}, \bibinfo {author}
  {\bibfnamefont {B.}~\bibnamefont {Fauqu{\'e}}}, \bibinfo {author}
  {\bibfnamefont {Y.}~\bibnamefont {Sidis}}, \bibinfo {author} {\bibfnamefont
  {G.}~\bibnamefont {Yu}}, \bibinfo {author} {\bibfnamefont {X.}~\bibnamefont
  {Zhao}}, \bibinfo {author} {\bibfnamefont {P.}~\bibnamefont {Bourges}},\ and\
  \bibinfo {author} {\bibfnamefont {M.}~\bibnamefont {Greven}},\ }\bibfield
  {title} {\bibinfo {title} {Unusual magnetic order in the pseudogap region of
  the superconductor \ce{HgBa2CuO_{4+\delta}}},\ }\href
  {https://doi.org/10.1038/nature07251} {\bibfield  {journal} {\bibinfo
  {journal} {Nature}\ }\textbf {\bibinfo {volume} {455}},\ \bibinfo {pages}
  {372} (\bibinfo {year} {2008})}\BibitemShut {NoStop}%
\bibitem [{\citenamefont {Li}\ \emph {et~al.}(2010)\citenamefont {Li},
  \citenamefont {Bal{\'e}dent}, \citenamefont {Yu}, \citenamefont
  {Bari{\v{s}}i{\'{c}}}, \citenamefont {Hradil}, \citenamefont {Mole},
  \citenamefont {Sidis}, \citenamefont {Steffens}, \citenamefont {Zhao},
  \citenamefont {Bourges},\ and\ \citenamefont {Greven}}]{Greven2010}%
  \BibitemOpen
  \bibfield  {author} {\bibinfo {author} {\bibfnamefont {Y.}~\bibnamefont
  {Li}}, \bibinfo {author} {\bibfnamefont {V.}~\bibnamefont {Bal{\'e}dent}},
  \bibinfo {author} {\bibfnamefont {G.}~\bibnamefont {Yu}}, \bibinfo {author}
  {\bibfnamefont {N.}~\bibnamefont {Bari{\v{s}}i{\'{c}}}}, \bibinfo {author}
  {\bibfnamefont {K.}~\bibnamefont {Hradil}}, \bibinfo {author} {\bibfnamefont
  {R.~A.}\ \bibnamefont {Mole}}, \bibinfo {author} {\bibfnamefont
  {Y.}~\bibnamefont {Sidis}}, \bibinfo {author} {\bibfnamefont
  {P.}~\bibnamefont {Steffens}}, \bibinfo {author} {\bibfnamefont
  {X.}~\bibnamefont {Zhao}}, \bibinfo {author} {\bibfnamefont {P.}~\bibnamefont
  {Bourges}},\ and\ \bibinfo {author} {\bibfnamefont {M.}~\bibnamefont
  {Greven}},\ }\bibfield  {title} {\bibinfo {title} {Hidden magnetic excitation
  in the pseudogap phase of a high-${T}_{c}$ superconductor},\ }\href
  {https://doi.org/10.1038/nature09477} {\bibfield  {journal} {\bibinfo
  {journal} {Nature}\ }\textbf {\bibinfo {volume} {468}},\ \bibinfo {pages}
  {283} (\bibinfo {year} {2010})}\BibitemShut {NoStop}%
\bibitem [{\citenamefont {Bourges}\ \emph {et~al.}(2021)\citenamefont
  {Bourges}, \citenamefont {Bounoua},\ and\ \citenamefont
  {Sidis}}]{Bourges2021}%
  \BibitemOpen
  \bibfield  {author} {\bibinfo {author} {\bibfnamefont {P.}~\bibnamefont
  {Bourges}}, \bibinfo {author} {\bibfnamefont {D.}~\bibnamefont {Bounoua}},\
  and\ \bibinfo {author} {\bibfnamefont {Y.}~\bibnamefont {Sidis}},\ }\bibfield
   {title} {\bibinfo {title} {Loop currents in quantum matter},\ }\href
  {https://doi.org/10.5802/crphys.84} {\bibfield  {journal} {\bibinfo
  {journal} {C. R. Physique}\ }\textbf {\bibinfo {volume} {22}},\ \bibinfo
  {pages} {7} (\bibinfo {year} {2021})}\BibitemShut {NoStop}%
\bibitem [{\citenamefont {Croft}\ \emph {et~al.}(2017)\citenamefont {Croft},
  \citenamefont {Blackburn}, \citenamefont {Kulda}, \citenamefont {Liang},
  \citenamefont {Bonn}, \citenamefont {Hardy},\ and\ \citenamefont
  {Hayden}}]{Croft2017}%
  \BibitemOpen
  \bibfield  {author} {\bibinfo {author} {\bibfnamefont {T.~P.}\ \bibnamefont
  {Croft}}, \bibinfo {author} {\bibfnamefont {E.}~\bibnamefont {Blackburn}},
  \bibinfo {author} {\bibfnamefont {J.}~\bibnamefont {Kulda}}, \bibinfo
  {author} {\bibfnamefont {R.}~\bibnamefont {Liang}}, \bibinfo {author}
  {\bibfnamefont {D.~A.}\ \bibnamefont {Bonn}}, \bibinfo {author}
  {\bibfnamefont {W.~N.}\ \bibnamefont {Hardy}},\ and\ \bibinfo {author}
  {\bibfnamefont {S.~M.}\ \bibnamefont {Hayden}},\ }\bibfield  {title}
  {\bibinfo {title} {No evidence for orbital loop currents in charge-ordered
  \ce{YBa2Cu3O_{6+x}} from polarized neutron diffraction},\ }\href
  {https://doi.org/10.1103/PhysRevB.96.214504} {\bibfield  {journal} {\bibinfo
  {journal} {Phys. Rev. B}\ }\textbf {\bibinfo {volume} {96}},\ \bibinfo
  {pages} {214504} (\bibinfo {year} {2017})}\BibitemShut {NoStop}%
\bibitem [{\citenamefont {Zhao}\ \emph {et~al.}(2016)\citenamefont {Zhao},
  \citenamefont {Torchinsky}, \citenamefont {Chu}, \citenamefont {Ivanov},
  \citenamefont {Lifshitz}, \citenamefont {Flint}, \citenamefont {Qi},
  \citenamefont {Cao},\ and\ \citenamefont {Hsieh}}]{Zhao2016}%
  \BibitemOpen
  \bibfield  {author} {\bibinfo {author} {\bibfnamefont {L.}~\bibnamefont
  {Zhao}}, \bibinfo {author} {\bibfnamefont {D.~H.}\ \bibnamefont
  {Torchinsky}}, \bibinfo {author} {\bibfnamefont {H.}~\bibnamefont {Chu}},
  \bibinfo {author} {\bibfnamefont {V.}~\bibnamefont {Ivanov}}, \bibinfo
  {author} {\bibfnamefont {R.}~\bibnamefont {Lifshitz}}, \bibinfo {author}
  {\bibfnamefont {R.}~\bibnamefont {Flint}}, \bibinfo {author} {\bibfnamefont
  {T.}~\bibnamefont {Qi}}, \bibinfo {author} {\bibfnamefont {G.}~\bibnamefont
  {Cao}},\ and\ \bibinfo {author} {\bibfnamefont {D.}~\bibnamefont {Hsieh}},\
  }\bibfield  {title} {\bibinfo {title} {Evidence of an odd-parity hidden order
  in a spin--orbit coupled correlated iridate},\ }\href
  {https://doi.org/10.1038/nphys3517} {\bibfield  {journal} {\bibinfo
  {journal} {Nat. Phys.}\ }\textbf {\bibinfo {volume} {12}},\ \bibinfo {pages}
  {32} (\bibinfo {year} {2016})}\BibitemShut {NoStop}%
\bibitem [{\citenamefont {Yan}\ \emph {et~al.}(2015)\citenamefont {Yan},
  \citenamefont {Ren}, \citenamefont {Xu}, \citenamefont {Xie}, \citenamefont
  {Tao}, \citenamefont {Choi}, \citenamefont {Lee}, \citenamefont {Choi},
  \citenamefont {Zhang},\ and\ \citenamefont {Feng}}]{Yan2015}%
  \BibitemOpen
  \bibfield  {author} {\bibinfo {author} {\bibfnamefont {Y.~J.}\ \bibnamefont
  {Yan}}, \bibinfo {author} {\bibfnamefont {M.~Q.}\ \bibnamefont {Ren}},
  \bibinfo {author} {\bibfnamefont {H.~C.}\ \bibnamefont {Xu}}, \bibinfo
  {author} {\bibfnamefont {B.~P.}\ \bibnamefont {Xie}}, \bibinfo {author}
  {\bibfnamefont {R.}~\bibnamefont {Tao}}, \bibinfo {author} {\bibfnamefont
  {H.~Y.}\ \bibnamefont {Choi}}, \bibinfo {author} {\bibfnamefont
  {N.}~\bibnamefont {Lee}}, \bibinfo {author} {\bibfnamefont {Y.~J.}\
  \bibnamefont {Choi}}, \bibinfo {author} {\bibfnamefont {T.}~\bibnamefont
  {Zhang}},\ and\ \bibinfo {author} {\bibfnamefont {D.~L.}\ \bibnamefont
  {Feng}},\ }\bibfield  {title} {\bibinfo {title} {Electron-doped \ce{Sr2IrO4}:
  An analogue of hole-doped cuprate superconductors demonstrated by scanning
  tunneling microscopy},\ }\href {https://doi.org/10.1103/PhysRevX.5.041018}
  {\bibfield  {journal} {\bibinfo  {journal} {Phys. Rev. X}\ }\textbf {\bibinfo
  {volume} {5}},\ \bibinfo {pages} {041018} (\bibinfo {year}
  {2015})}\BibitemShut {NoStop}%
\bibitem [{\citenamefont {Kim}\ \emph {et~al.}(2016)\citenamefont {Kim},
  \citenamefont {Sung}, \citenamefont {Denlinger},\ and\ \citenamefont
  {Kim}}]{Kim2016}%
  \BibitemOpen
  \bibfield  {author} {\bibinfo {author} {\bibfnamefont {Y.~K.}\ \bibnamefont
  {Kim}}, \bibinfo {author} {\bibfnamefont {N.~H.}\ \bibnamefont {Sung}},
  \bibinfo {author} {\bibfnamefont {J.~D.}\ \bibnamefont {Denlinger}},\ and\
  \bibinfo {author} {\bibfnamefont {B.~J.}\ \bibnamefont {Kim}},\ }\bibfield
  {title} {\bibinfo {title} {Observation of a ${d}$-wave gap in electron-doped
  \ce{Sr2IrO4}},\ }\href {https://doi.org/10.1038/nphys3503} {\bibfield
  {journal} {\bibinfo  {journal} {Nat. Phys.}\ }\textbf {\bibinfo {volume}
  {12}},\ \bibinfo {pages} {37} (\bibinfo {year} {2016})}\BibitemShut {NoStop}%
\bibitem [{\citenamefont {Mielke}\ \emph {et~al.}(2022)\citenamefont {Mielke},
  \citenamefont {Das}, \citenamefont {Yin}, \citenamefont {Liu}, \citenamefont
  {Gupta}, \citenamefont {Jiang}, \citenamefont {Medarde}, \citenamefont {Wu},
  \citenamefont {Lei}, \citenamefont {Chang}, \citenamefont {Dai},
  \citenamefont {Si}, \citenamefont {Miao}, \citenamefont {Thomale},
  \citenamefont {Neupert}, \citenamefont {Shi}, \citenamefont {Khasanov},
  \citenamefont {Hasan}, \citenamefont {Luetkens},\ and\ \citenamefont
  {Guguchia}}]{Mielke2022}%
  \BibitemOpen
  \bibfield  {author} {\bibinfo {author} {\bibfnamefont {C.}~\bibnamefont
  {Mielke}}, \bibinfo {author} {\bibfnamefont {D.}~\bibnamefont {Das}},
  \bibinfo {author} {\bibfnamefont {J.-X.}\ \bibnamefont {Yin}}, \bibinfo
  {author} {\bibfnamefont {H.}~\bibnamefont {Liu}}, \bibinfo {author}
  {\bibfnamefont {R.}~\bibnamefont {Gupta}}, \bibinfo {author} {\bibfnamefont
  {Y.-X.}\ \bibnamefont {Jiang}}, \bibinfo {author} {\bibfnamefont
  {M.}~\bibnamefont {Medarde}}, \bibinfo {author} {\bibfnamefont
  {X.}~\bibnamefont {Wu}}, \bibinfo {author} {\bibfnamefont {H.~C.}\
  \bibnamefont {Lei}}, \bibinfo {author} {\bibfnamefont {J.}~\bibnamefont
  {Chang}}, \bibinfo {author} {\bibfnamefont {P.}~\bibnamefont {Dai}}, \bibinfo
  {author} {\bibfnamefont {Q.}~\bibnamefont {Si}}, \bibinfo {author}
  {\bibfnamefont {H.}~\bibnamefont {Miao}}, \bibinfo {author} {\bibfnamefont
  {R.}~\bibnamefont {Thomale}}, \bibinfo {author} {\bibfnamefont
  {T.}~\bibnamefont {Neupert}}, \bibinfo {author} {\bibfnamefont
  {Y.}~\bibnamefont {Shi}}, \bibinfo {author} {\bibfnamefont {R.}~\bibnamefont
  {Khasanov}}, \bibinfo {author} {\bibfnamefont {M.~Z.}\ \bibnamefont {Hasan}},
  \bibinfo {author} {\bibfnamefont {H.}~\bibnamefont {Luetkens}},\ and\
  \bibinfo {author} {\bibfnamefont {Z.}~\bibnamefont {Guguchia}},\ }\bibfield
  {title} {\bibinfo {title} {Time-reversal symmetry-breaking charge order in a
  kagome superconductor},\ }\href {https://doi.org/10.1038/s41586-021-04327-z}
  {\bibfield  {journal} {\bibinfo  {journal} {Nature}\ }\textbf {\bibinfo
  {volume} {602}},\ \bibinfo {pages} {245} (\bibinfo {year}
  {2022})}\BibitemShut {NoStop}%
\bibitem [{\citenamefont {Sun}\ and\ \citenamefont {Fradkin}(2008)}]{Sun2008}%
  \BibitemOpen
  \bibfield  {author} {\bibinfo {author} {\bibfnamefont {K.}~\bibnamefont
  {Sun}}\ and\ \bibinfo {author} {\bibfnamefont {E.}~\bibnamefont {Fradkin}},\
  }\bibfield  {title} {\bibinfo {title} {Time-reversal symmetry breaking and
  spontaneous anomalous {H}all effect in {F}ermi fluids},\ }\href
  {https://doi.org/10.1103/PhysRevB.78.245122} {\bibfield  {journal} {\bibinfo
  {journal} {Phys. Rev. B}\ }\textbf {\bibinfo {volume} {78}},\ \bibinfo
  {pages} {245122} (\bibinfo {year} {2008})}\BibitemShut {NoStop}%
\bibitem [{\citenamefont {Castro}\ \emph {et~al.}(2011)\citenamefont {Castro},
  \citenamefont {Grushin}, \citenamefont {Valenzuela}, \citenamefont
  {Vozmediano}, \citenamefont {Cortijo},\ and\ \citenamefont
  {de~Juan}}]{Castro2011}%
  \BibitemOpen
  \bibfield  {author} {\bibinfo {author} {\bibfnamefont {E.~V.}\ \bibnamefont
  {Castro}}, \bibinfo {author} {\bibfnamefont {A.~G.}\ \bibnamefont {Grushin}},
  \bibinfo {author} {\bibfnamefont {B.}~\bibnamefont {Valenzuela}}, \bibinfo
  {author} {\bibfnamefont {M.~A.~H.}\ \bibnamefont {Vozmediano}}, \bibinfo
  {author} {\bibfnamefont {A.}~\bibnamefont {Cortijo}},\ and\ \bibinfo {author}
  {\bibfnamefont {F.}~\bibnamefont {de~Juan}},\ }\bibfield  {title} {\bibinfo
  {title} {Topological {F}ermi liquids from {C}oulomb interactions in the doped
  honeycomb lattice},\ }\href {https://doi.org/10.1103/PhysRevLett.107.106402}
  {\bibfield  {journal} {\bibinfo  {journal} {Phys. Rev. Lett.}\ }\textbf
  {\bibinfo {volume} {107}},\ \bibinfo {pages} {106402} (\bibinfo {year}
  {2011})}\BibitemShut {NoStop}%
\bibitem [{\citenamefont {Sur}\ \emph {et~al.}(2018)\citenamefont {Sur},
  \citenamefont {Gong}, \citenamefont {Yang},\ and\ \citenamefont
  {Vafek}}]{Sur2018}%
  \BibitemOpen
  \bibfield  {author} {\bibinfo {author} {\bibfnamefont {S.}~\bibnamefont
  {Sur}}, \bibinfo {author} {\bibfnamefont {S.-S.}\ \bibnamefont {Gong}},
  \bibinfo {author} {\bibfnamefont {K.}~\bibnamefont {Yang}},\ and\ \bibinfo
  {author} {\bibfnamefont {O.}~\bibnamefont {Vafek}},\ }\bibfield  {title}
  {\bibinfo {title} {Quantum anomalous {H}all insulator stabilized by competing
  interactions},\ }\href {https://doi.org/10.1103/PhysRevB.98.125144}
  {\bibfield  {journal} {\bibinfo  {journal} {Phys. Rev. B}\ }\textbf {\bibinfo
  {volume} {98}},\ \bibinfo {pages} {125144} (\bibinfo {year}
  {2018})}\BibitemShut {NoStop}%
\bibitem [{\citenamefont {Scheurer}\ and\ \citenamefont
  {Sachdev}(2018)}]{Scheurer2018}%
  \BibitemOpen
  \bibfield  {author} {\bibinfo {author} {\bibfnamefont {M.~S.}\ \bibnamefont
  {Scheurer}}\ and\ \bibinfo {author} {\bibfnamefont {S.}~\bibnamefont
  {Sachdev}},\ }\bibfield  {title} {\bibinfo {title} {Orbital currents in
  insulating and doped antiferromagnets},\ }\href
  {https://doi.org/10.1103/PhysRevB.98.235126} {\bibfield  {journal} {\bibinfo
  {journal} {Phys. Rev. B}\ }\textbf {\bibinfo {volume} {98}},\ \bibinfo
  {pages} {235126} (\bibinfo {year} {2018})}\BibitemShut {NoStop}%
\bibitem [{\citenamefont {Lederer}\ \emph {et~al.}(2015)\citenamefont
  {Lederer}, \citenamefont {Schattner}, \citenamefont {Berg},\ and\
  \citenamefont {Kivelson}}]{Lederer2015}%
  \BibitemOpen
  \bibfield  {author} {\bibinfo {author} {\bibfnamefont {S.}~\bibnamefont
  {Lederer}}, \bibinfo {author} {\bibfnamefont {Y.}~\bibnamefont {Schattner}},
  \bibinfo {author} {\bibfnamefont {E.}~\bibnamefont {Berg}},\ and\ \bibinfo
  {author} {\bibfnamefont {S.~A.}\ \bibnamefont {Kivelson}},\ }\bibfield
  {title} {\bibinfo {title} {Enhancement of superconductivity near a nematic
  quantum critical point},\ }\href
  {https://doi.org/10.1103/PhysRevLett.114.097001} {\bibfield  {journal}
  {\bibinfo  {journal} {Phys. Rev. Lett.}\ }\textbf {\bibinfo {volume} {114}},\
  \bibinfo {pages} {097001} (\bibinfo {year} {2015})}\BibitemShut {NoStop}%
\bibitem [{\citenamefont {Lederer}\ \emph {et~al.}(2017)\citenamefont
  {Lederer}, \citenamefont {Schattner}, \citenamefont {Berg},\ and\
  \citenamefont {Kivelson}}]{Lederer2017}%
  \BibitemOpen
  \bibfield  {author} {\bibinfo {author} {\bibfnamefont {S.}~\bibnamefont
  {Lederer}}, \bibinfo {author} {\bibfnamefont {Y.}~\bibnamefont {Schattner}},
  \bibinfo {author} {\bibfnamefont {E.}~\bibnamefont {Berg}},\ and\ \bibinfo
  {author} {\bibfnamefont {S.~A.}\ \bibnamefont {Kivelson}},\ }\bibfield
  {title} {\bibinfo {title} {Superconductivity and non-{F}ermi liquid behavior
  near a nematic quantum critical point},\ }\href
  {https://doi.org/10.1073/pnas.1620651114} {\bibfield  {journal} {\bibinfo
  {journal} {PNAS}\ }\textbf {\bibinfo {volume} {114}},\ \bibinfo {pages}
  {4905} (\bibinfo {year} {2017})}\BibitemShut {NoStop}%
\bibitem [{\citenamefont {Klein}\ and\ \citenamefont
  {Chubukov}(2018)}]{Klein2018}%
  \BibitemOpen
  \bibfield  {author} {\bibinfo {author} {\bibfnamefont {A.}~\bibnamefont
  {Klein}}\ and\ \bibinfo {author} {\bibfnamefont {A.}~\bibnamefont
  {Chubukov}},\ }\bibfield  {title} {\bibinfo {title} {Superconductivity near a
  nematic quantum critical point: Interplay between hot and lukewarm regions},\
  }\href {https://doi.org/10.1103/PhysRevB.98.220501} {\bibfield  {journal}
  {\bibinfo  {journal} {Phys. Rev. B}\ }\textbf {\bibinfo {volume} {98}},\
  \bibinfo {pages} {220501} (\bibinfo {year} {2018})}\BibitemShut {NoStop}%
\bibitem [{\citenamefont {Kozii}\ and\ \citenamefont {Fu}(2015)}]{Kozii2015}%
  \BibitemOpen
  \bibfield  {author} {\bibinfo {author} {\bibfnamefont {V.}~\bibnamefont
  {Kozii}}\ and\ \bibinfo {author} {\bibfnamefont {L.}~\bibnamefont {Fu}},\
  }\bibfield  {title} {\bibinfo {title} {Odd-parity superconductivity in the
  vicinity of inversion symmetry breaking in spin-orbit-coupled systems},\
  }\href {https://doi.org/10.1103/PhysRevLett.115.207002} {\bibfield  {journal}
  {\bibinfo  {journal} {Phys. Rev. Lett.}\ }\textbf {\bibinfo {volume} {115}},\
  \bibinfo {pages} {207002} (\bibinfo {year} {2015})}\BibitemShut {NoStop}%
\bibitem [{\citenamefont {Klein}\ \emph {et~al.}(2023)\citenamefont {Klein},
  \citenamefont {Kozii}, \citenamefont {Ruhman},\ and\ \citenamefont
  {Fernandes}}]{Klein2023}%
  \BibitemOpen
  \bibfield  {author} {\bibinfo {author} {\bibfnamefont {A.}~\bibnamefont
  {Klein}}, \bibinfo {author} {\bibfnamefont {V.}~\bibnamefont {Kozii}},
  \bibinfo {author} {\bibfnamefont {J.}~\bibnamefont {Ruhman}},\ and\ \bibinfo
  {author} {\bibfnamefont {R.~M.}\ \bibnamefont {Fernandes}},\ }\bibfield
  {title} {\bibinfo {title} {Theory of criticality for quantum ferroelectric
  metals},\ }\href {https://doi.org/10.1103/PhysRevB.107.165110} {\bibfield
  {journal} {\bibinfo  {journal} {Phys. Rev. B}\ }\textbf {\bibinfo {volume}
  {107}},\ \bibinfo {pages} {165110} (\bibinfo {year} {2023})}\BibitemShut
  {NoStop}%
\bibitem [{\citenamefont {Roussev}\ and\ \citenamefont
  {Millis}(2001)}]{Millis2001}%
  \BibitemOpen
  \bibfield  {author} {\bibinfo {author} {\bibfnamefont {R.}~\bibnamefont
  {Roussev}}\ and\ \bibinfo {author} {\bibfnamefont {A.~J.}\ \bibnamefont
  {Millis}},\ }\bibfield  {title} {\bibinfo {title} {Quantum critical effects
  on transition temperature of magnetically mediated ${p}$-wave
  superconductivity},\ }\href {https://doi.org/10.1103/PhysRevB.63.140504}
  {\bibfield  {journal} {\bibinfo  {journal} {Phys. Rev. B}\ }\textbf {\bibinfo
  {volume} {63}},\ \bibinfo {pages} {140504} (\bibinfo {year}
  {2001})}\BibitemShut {NoStop}%
\bibitem [{\citenamefont {Chubukov}\ \emph {et~al.}(2003)\citenamefont
  {Chubukov}, \citenamefont {Finkel’stein}, \citenamefont {Haslinger},\ and\
  \citenamefont {Morr}}]{Chubukov2003}%
  \BibitemOpen
  \bibfield  {author} {\bibinfo {author} {\bibfnamefont {A.~V.}\ \bibnamefont
  {Chubukov}}, \bibinfo {author} {\bibfnamefont {A.~M.}\ \bibnamefont
  {Finkel’stein}}, \bibinfo {author} {\bibfnamefont {R.}~\bibnamefont
  {Haslinger}},\ and\ \bibinfo {author} {\bibfnamefont {D.~K.}\ \bibnamefont
  {Morr}},\ }\bibfield  {title} {\bibinfo {title} {First-order superconducting
  transition near a ferromagnetic quantum critical point},\ }\href
  {https://doi.org/10.1103/PhysRevLett.90.077002} {\bibfield  {journal}
  {\bibinfo  {journal} {Phys. Rev. Lett.}\ }\textbf {\bibinfo {volume} {90}},\
  \bibinfo {pages} {077002} (\bibinfo {year} {2003})}\BibitemShut {NoStop}%
\bibitem [{\citenamefont {Abanov}\ \emph {et~al.}(2001)\citenamefont {Abanov},
  \citenamefont {Chubukov},\ and\ \citenamefont {Schmalian}}]{Abanov2001}%
  \BibitemOpen
  \bibfield  {author} {\bibinfo {author} {\bibfnamefont {A.}~\bibnamefont
  {Abanov}}, \bibinfo {author} {\bibfnamefont {A.~V.}\ \bibnamefont
  {Chubukov}},\ and\ \bibinfo {author} {\bibfnamefont {J.}~\bibnamefont
  {Schmalian}},\ }\bibfield  {title} {\bibinfo {title} {Quantum-critical
  superconductivity in underdoped cuprates},\ }\href
  {https://doi.org/10.1209/epl/i2001-00425-9} {\bibfield  {journal} {\bibinfo
  {journal} {Europhys. Lett.}\ }\textbf {\bibinfo {volume} {55}},\ \bibinfo
  {pages} {369} (\bibinfo {year} {2001})}\BibitemShut {NoStop}%
\bibitem [{\citenamefont {Varma}(2012)}]{Varma2012}%
  \BibitemOpen
  \bibfield  {author} {\bibinfo {author} {\bibfnamefont {C.~M.}\ \bibnamefont
  {Varma}},\ }\bibfield  {title} {\bibinfo {title} {Considerations on the
  mechanisms and transition temperatures of superconductivity induced by
  electronic fluctuations},\ }\href
  {https://doi.org/10.1088/0034-4885/75/5/052501} {\bibfield  {journal}
  {\bibinfo  {journal} {Rep. Prog. Phys.}\ }\textbf {\bibinfo {volume} {75}},\
  \bibinfo {pages} {052501} (\bibinfo {year} {2012})}\BibitemShut {NoStop}%
\bibitem [{\citenamefont {Wang}\ \emph {et~al.}(2017)\citenamefont {Wang},
  \citenamefont {Schattner}, \citenamefont {Berg},\ and\ \citenamefont
  {Fernandes}}]{Wang2017}%
  \BibitemOpen
  \bibfield  {author} {\bibinfo {author} {\bibfnamefont {X.}~\bibnamefont
  {Wang}}, \bibinfo {author} {\bibfnamefont {Y.}~\bibnamefont {Schattner}},
  \bibinfo {author} {\bibfnamefont {E.}~\bibnamefont {Berg}},\ and\ \bibinfo
  {author} {\bibfnamefont {R.~M.}\ \bibnamefont {Fernandes}},\ }\bibfield
  {title} {\bibinfo {title} {Superconductivity mediated by quantum critical
  antiferromagnetic fluctuations: The rise and fall of hot spots},\ }\href
  {https://doi.org/10.1103/PhysRevB.95.174520} {\bibfield  {journal} {\bibinfo
  {journal} {Phys. Rev. B}\ }\textbf {\bibinfo {volume} {95}},\ \bibinfo
  {pages} {174520} (\bibinfo {year} {2017})}\BibitemShut {NoStop}%
\bibitem [{\citenamefont {Xu}\ \emph {et~al.}(2017)\citenamefont {Xu},
  \citenamefont {Sun}, \citenamefont {Schattner}, \citenamefont {Berg},\ and\
  \citenamefont {Meng}}]{Xu2017}%
  \BibitemOpen
  \bibfield  {author} {\bibinfo {author} {\bibfnamefont {X.~Y.}\ \bibnamefont
  {Xu}}, \bibinfo {author} {\bibfnamefont {K.}~\bibnamefont {Sun}}, \bibinfo
  {author} {\bibfnamefont {Y.}~\bibnamefont {Schattner}}, \bibinfo {author}
  {\bibfnamefont {E.}~\bibnamefont {Berg}},\ and\ \bibinfo {author}
  {\bibfnamefont {Z.~Y.}\ \bibnamefont {Meng}},\ }\bibfield  {title} {\bibinfo
  {title} {Non-{F}ermi liquid at {$(2+1)-\mathrm{D}$} ferromagnetic quantum
  critical point},\ }\href {https://doi.org/10.1103/PhysRevX.7.031058}
  {\bibfield  {journal} {\bibinfo  {journal} {Phys. Rev. X}\ }\textbf {\bibinfo
  {volume} {7}},\ \bibinfo {pages} {031058} (\bibinfo {year}
  {2017})}\BibitemShut {NoStop}%
\bibitem [{\citenamefont {Xu}\ \emph {et~al.}(2020)\citenamefont {Xu},
  \citenamefont {Klein}, \citenamefont {Sun}, \citenamefont {Chubukov},\ and\
  \citenamefont {Meng}}]{Xu2020}%
  \BibitemOpen
  \bibfield  {author} {\bibinfo {author} {\bibfnamefont {X.~Y.}\ \bibnamefont
  {Xu}}, \bibinfo {author} {\bibfnamefont {A.}~\bibnamefont {Klein}}, \bibinfo
  {author} {\bibfnamefont {K.}~\bibnamefont {Sun}}, \bibinfo {author}
  {\bibfnamefont {A.~V.}\ \bibnamefont {Chubukov}},\ and\ \bibinfo {author}
  {\bibfnamefont {Z.~Y.}\ \bibnamefont {Meng}},\ }\bibfield  {title} {\bibinfo
  {title} {Identification of non-{F}ermi liquid fermionic self-energy from
  quantum {M}onte {C}arlo data},\ }\href
  {https://doi.org/10.1038/s41535-020-00266-6} {\bibfield  {journal} {\bibinfo
  {journal} {npj Quantum Mater.}\ }\textbf {\bibinfo {volume} {5}},\ \bibinfo
  {pages} {65} (\bibinfo {year} {2020})}\BibitemShut {NoStop}%
\bibitem [{\citenamefont {Abanov}\ and\ \citenamefont
  {Chubukov}(2020)}]{ChubukovI_2020}%
  \BibitemOpen
  \bibfield  {author} {\bibinfo {author} {\bibfnamefont {A.}~\bibnamefont
  {Abanov}}\ and\ \bibinfo {author} {\bibfnamefont {A.~V.}\ \bibnamefont
  {Chubukov}},\ }\bibfield  {title} {\bibinfo {title} {Interplay between
  superconductivity and non-{F}ermi liquid at a quantum critical point in a
  metal. i. the $\ensuremath{\gamma}$ model and its phase diagram at ${T}=0$:
  The case $0<\gamma<1$},\ }\href {https://doi.org/10.1103/PhysRevB.102.024524}
  {\bibfield  {journal} {\bibinfo  {journal} {Phys. Rev. B}\ }\textbf {\bibinfo
  {volume} {102}},\ \bibinfo {pages} {024524} (\bibinfo {year}
  {2020})}\BibitemShut {NoStop}%
\bibitem [{\citenamefont {Wu}\ \emph {et~al.}(2020)\citenamefont {Wu},
  \citenamefont {Abanov}, \citenamefont {Wang},\ and\ \citenamefont
  {Chubukov}}]{ChubukovII_2020}%
  \BibitemOpen
  \bibfield  {author} {\bibinfo {author} {\bibfnamefont {Y.-M.}\ \bibnamefont
  {Wu}}, \bibinfo {author} {\bibfnamefont {A.}~\bibnamefont {Abanov}}, \bibinfo
  {author} {\bibfnamefont {Y.}~\bibnamefont {Wang}},\ and\ \bibinfo {author}
  {\bibfnamefont {A.~V.}\ \bibnamefont {Chubukov}},\ }\bibfield  {title}
  {\bibinfo {title} {Interplay between superconductivity and non-{F}ermi liquid
  at a quantum critical point in a metal. ii. the $\ensuremath{\gamma}$ model
  at a finite ${T}$ for $0<\gamma<1$},\ }\href
  {https://doi.org/10.1103/PhysRevB.102.024525} {\bibfield  {journal} {\bibinfo
   {journal} {Phys. Rev. B}\ }\textbf {\bibinfo {volume} {102}},\ \bibinfo
  {pages} {024525} (\bibinfo {year} {2020})}\BibitemShut {NoStop}%
\bibitem [{\citenamefont {Proust}\ \emph {et~al.}(2002)\citenamefont {Proust},
  \citenamefont {Boaknin}, \citenamefont {Hill}, \citenamefont {Taillefer},\
  and\ \citenamefont {Mackenzie}}]{Proust2002}%
  \BibitemOpen
  \bibfield  {author} {\bibinfo {author} {\bibfnamefont {C.}~\bibnamefont
  {Proust}}, \bibinfo {author} {\bibfnamefont {E.}~\bibnamefont {Boaknin}},
  \bibinfo {author} {\bibfnamefont {R.~W.}\ \bibnamefont {Hill}}, \bibinfo
  {author} {\bibfnamefont {L.}~\bibnamefont {Taillefer}},\ and\ \bibinfo
  {author} {\bibfnamefont {A.~P.}\ \bibnamefont {Mackenzie}},\ }\bibfield
  {title} {\bibinfo {title} {Heat transport in a strongly overdoped cuprate:
  {F}ermi liquid and a pure ${d}$-wave {BCS} superconductor},\ }\href
  {https://doi.org/10.1103/PhysRevLett.89.147003} {\bibfield  {journal}
  {\bibinfo  {journal} {Phys. Rev. Lett.}\ }\textbf {\bibinfo {volume} {89}},\
  \bibinfo {pages} {147003} (\bibinfo {year} {2002})}\BibitemShut {NoStop}%
\bibitem [{\citenamefont {Nakamae}\ \emph {et~al.}(2003)\citenamefont
  {Nakamae}, \citenamefont {Behnia}, \citenamefont {Mangkorntong},
  \citenamefont {Nohara}, \citenamefont {Takagi}, \citenamefont {Yates},\ and\
  \citenamefont {Hussey}}]{Nakamae2003}%
  \BibitemOpen
  \bibfield  {author} {\bibinfo {author} {\bibfnamefont {S.}~\bibnamefont
  {Nakamae}}, \bibinfo {author} {\bibfnamefont {K.}~\bibnamefont {Behnia}},
  \bibinfo {author} {\bibfnamefont {N.}~\bibnamefont {Mangkorntong}}, \bibinfo
  {author} {\bibfnamefont {M.}~\bibnamefont {Nohara}}, \bibinfo {author}
  {\bibfnamefont {H.}~\bibnamefont {Takagi}}, \bibinfo {author} {\bibfnamefont
  {S.~J.~C.}\ \bibnamefont {Yates}},\ and\ \bibinfo {author} {\bibfnamefont
  {N.~E.}\ \bibnamefont {Hussey}},\ }\bibfield  {title} {\bibinfo {title}
  {Electronic ground state of heavily overdoped nonsuperconducting
  \ce{La_{2-x}Sr_xCuO4}},\ }\href {https://doi.org/10.1103/PhysRevB.68.100502}
  {\bibfield  {journal} {\bibinfo  {journal} {Phys. Rev. B}\ }\textbf {\bibinfo
  {volume} {68}},\ \bibinfo {pages} {100502} (\bibinfo {year}
  {2003})}\BibitemShut {NoStop}%
\bibitem [{\citenamefont {Hussey}\ \emph {et~al.}(2003)\citenamefont {Hussey},
  \citenamefont {Abdel-Jawad}, \citenamefont {Carrington}, \citenamefont
  {Mackenzie},\ and\ \citenamefont {Balicas}}]{Hussey2003}%
  \BibitemOpen
  \bibfield  {author} {\bibinfo {author} {\bibfnamefont {N.~E.}\ \bibnamefont
  {Hussey}}, \bibinfo {author} {\bibfnamefont {M.}~\bibnamefont {Abdel-Jawad}},
  \bibinfo {author} {\bibfnamefont {A.}~\bibnamefont {Carrington}}, \bibinfo
  {author} {\bibfnamefont {A.~P.}\ \bibnamefont {Mackenzie}},\ and\ \bibinfo
  {author} {\bibfnamefont {L.}~\bibnamefont {Balicas}},\ }\bibfield  {title}
  {\bibinfo {title} {A coherent three-dimensional {F}ermi surface in a
  high-transition-temperature superconductor},\ }\href
  {https://doi.org/10.1038/nature01981} {\bibfield  {journal} {\bibinfo
  {journal} {Nature}\ }\textbf {\bibinfo {volume} {425}},\ \bibinfo {pages}
  {814} (\bibinfo {year} {2003})}\BibitemShut {NoStop}%
\bibitem [{\citenamefont {Damascelli}\ \emph {et~al.}(2003)\citenamefont
  {Damascelli}, \citenamefont {Hussain},\ and\ \citenamefont
  {Shen}}]{Damascelli2003}%
  \BibitemOpen
  \bibfield  {author} {\bibinfo {author} {\bibfnamefont {A.}~\bibnamefont
  {Damascelli}}, \bibinfo {author} {\bibfnamefont {Z.}~\bibnamefont
  {Hussain}},\ and\ \bibinfo {author} {\bibfnamefont {Z.-X.}\ \bibnamefont
  {Shen}},\ }\bibfield  {title} {\bibinfo {title} {Angle-resolved photoemission
  studies of the cuprate superconductors},\ }\href
  {https://doi.org/10.1103/RevModPhys.75.473} {\bibfield  {journal} {\bibinfo
  {journal} {Rev. Mod. Phys.}\ }\textbf {\bibinfo {volume} {75}},\ \bibinfo
  {pages} {473} (\bibinfo {year} {2003})}\BibitemShut {NoStop}%
\bibitem [{\citenamefont {Plat\'{e}}\ \emph {et~al.}(2005)\citenamefont
  {Plat\'{e}}, \citenamefont {Mottershead}, \citenamefont {Elfimov},
  \citenamefont {Peets}, \citenamefont {Liang}, \citenamefont {Bonn},
  \citenamefont {Hardy}, \citenamefont {Chiuzbaian}, \citenamefont {Falub},
  \citenamefont {Shi}, \citenamefont {Patthey},\ and\ \citenamefont
  {Damascelli}}]{Plate2005}%
  \BibitemOpen
  \bibfield  {author} {\bibinfo {author} {\bibfnamefont {M.}~\bibnamefont
  {Plat\'{e}}}, \bibinfo {author} {\bibfnamefont {J.~D.~F.}\ \bibnamefont
  {Mottershead}}, \bibinfo {author} {\bibfnamefont {I.~S.}\ \bibnamefont
  {Elfimov}}, \bibinfo {author} {\bibfnamefont {D.~C.}\ \bibnamefont {Peets}},
  \bibinfo {author} {\bibfnamefont {R.}~\bibnamefont {Liang}}, \bibinfo
  {author} {\bibfnamefont {D.~A.}\ \bibnamefont {Bonn}}, \bibinfo {author}
  {\bibfnamefont {W.~N.}\ \bibnamefont {Hardy}}, \bibinfo {author}
  {\bibfnamefont {S.}~\bibnamefont {Chiuzbaian}}, \bibinfo {author}
  {\bibfnamefont {M.}~\bibnamefont {Falub}}, \bibinfo {author} {\bibfnamefont
  {M.}~\bibnamefont {Shi}}, \bibinfo {author} {\bibfnamefont {L.}~\bibnamefont
  {Patthey}},\ and\ \bibinfo {author} {\bibfnamefont {A.}~\bibnamefont
  {Damascelli}},\ }\bibfield  {title} {\bibinfo {title} {{F}ermi surface and
  quasiparticle excitations of overdoped \ce{Tl2Ba2CuO_{6+\delta}}},\ }\href
  {https://doi.org/10.1103/PhysRevLett.95.077001} {\bibfield  {journal}
  {\bibinfo  {journal} {Phys. Rev. Lett.}\ }\textbf {\bibinfo {volume} {95}},\
  \bibinfo {pages} {077001} (\bibinfo {year} {2005})}\BibitemShut {NoStop}%
\bibitem [{\citenamefont {Horio}\ \emph {et~al.}(2018)\citenamefont {Horio},
  \citenamefont {Hauser}, \citenamefont {Sassa}, \citenamefont {Mingazheva},
  \citenamefont {Sutter}, \citenamefont {Kramer}, \citenamefont {Cook},
  \citenamefont {Nocerino}, \citenamefont {Forslund}, \citenamefont
  {Tjernberg}, \citenamefont {Kobayashi}, \citenamefont {Chikina},
  \citenamefont {Schr\"{o}ter}, \citenamefont {Krieger}, \citenamefont
  {Schmitt}, \citenamefont {Strocov}, \citenamefont {Pyon}, \citenamefont
  {Takayama}, \citenamefont {Takagi}, \citenamefont {Lipscombe}, \citenamefont
  {Hayden}, \citenamefont {Ishikado}, \citenamefont {Eisaki}, \citenamefont
  {Neupert}, \citenamefont {M\aa{}nsson}, \citenamefont {Matt},\ and\
  \citenamefont {Chang}}]{Horio2018}%
  \BibitemOpen
  \bibfield  {author} {\bibinfo {author} {\bibfnamefont {M.}~\bibnamefont
  {Horio}}, \bibinfo {author} {\bibfnamefont {K.}~\bibnamefont {Hauser}},
  \bibinfo {author} {\bibfnamefont {Y.}~\bibnamefont {Sassa}}, \bibinfo
  {author} {\bibfnamefont {Z.}~\bibnamefont {Mingazheva}}, \bibinfo {author}
  {\bibfnamefont {D.}~\bibnamefont {Sutter}}, \bibinfo {author} {\bibfnamefont
  {K.}~\bibnamefont {Kramer}}, \bibinfo {author} {\bibfnamefont
  {A.}~\bibnamefont {Cook}}, \bibinfo {author} {\bibfnamefont {E.}~\bibnamefont
  {Nocerino}}, \bibinfo {author} {\bibfnamefont {O.~K.}\ \bibnamefont
  {Forslund}}, \bibinfo {author} {\bibfnamefont {O.}~\bibnamefont {Tjernberg}},
  \bibinfo {author} {\bibfnamefont {M.}~\bibnamefont {Kobayashi}}, \bibinfo
  {author} {\bibfnamefont {A.}~\bibnamefont {Chikina}}, \bibinfo {author}
  {\bibfnamefont {N.~B.~M.}\ \bibnamefont {Schr\"{o}ter}}, \bibinfo {author}
  {\bibfnamefont {J.~A.}\ \bibnamefont {Krieger}}, \bibinfo {author}
  {\bibfnamefont {T.}~\bibnamefont {Schmitt}}, \bibinfo {author} {\bibfnamefont
  {V.~N.}\ \bibnamefont {Strocov}}, \bibinfo {author} {\bibfnamefont
  {S.}~\bibnamefont {Pyon}}, \bibinfo {author} {\bibfnamefont {T.}~\bibnamefont
  {Takayama}}, \bibinfo {author} {\bibfnamefont {H.}~\bibnamefont {Takagi}},
  \bibinfo {author} {\bibfnamefont {O.~J.}\ \bibnamefont {Lipscombe}}, \bibinfo
  {author} {\bibfnamefont {S.~M.}\ \bibnamefont {Hayden}}, \bibinfo {author}
  {\bibfnamefont {M.}~\bibnamefont {Ishikado}}, \bibinfo {author}
  {\bibfnamefont {H.}~\bibnamefont {Eisaki}}, \bibinfo {author} {\bibfnamefont
  {T.}~\bibnamefont {Neupert}}, \bibinfo {author} {\bibfnamefont
  {M.}~\bibnamefont {M\aa{}nsson}}, \bibinfo {author} {\bibfnamefont {C.~E.}\
  \bibnamefont {Matt}},\ and\ \bibinfo {author} {\bibfnamefont
  {J.}~\bibnamefont {Chang}},\ }\bibfield  {title} {\bibinfo {title}
  {Three-dimensional {F}ermi surface of overdoped \ce{La}-based cuprates},\
  }\href {https://doi.org/10.1103/PhysRevLett.121.077004} {\bibfield  {journal}
  {\bibinfo  {journal} {Phys. Rev. Lett.}\ }\textbf {\bibinfo {volume} {121}},\
  \bibinfo {pages} {077004} (\bibinfo {year} {2018})}\BibitemShut {NoStop}%
\bibitem [{\citenamefont {Vignolle}\ \emph {et~al.}(2008)\citenamefont
  {Vignolle}, \citenamefont {Carrington}, \citenamefont {Cooper}, \citenamefont
  {French}, \citenamefont {Mackenzie}, \citenamefont {Jaudet}, \citenamefont
  {Vignolles}, \citenamefont {Proust},\ and\ \citenamefont
  {Hussey}}]{Vignolle2008}%
  \BibitemOpen
  \bibfield  {author} {\bibinfo {author} {\bibfnamefont {B.}~\bibnamefont
  {Vignolle}}, \bibinfo {author} {\bibfnamefont {A.}~\bibnamefont
  {Carrington}}, \bibinfo {author} {\bibfnamefont {R.~A.}\ \bibnamefont
  {Cooper}}, \bibinfo {author} {\bibfnamefont {M.~M.~J.}\ \bibnamefont
  {French}}, \bibinfo {author} {\bibfnamefont {A.~P.}\ \bibnamefont
  {Mackenzie}}, \bibinfo {author} {\bibfnamefont {C.}~\bibnamefont {Jaudet}},
  \bibinfo {author} {\bibfnamefont {D.}~\bibnamefont {Vignolles}}, \bibinfo
  {author} {\bibfnamefont {C.}~\bibnamefont {Proust}},\ and\ \bibinfo {author}
  {\bibfnamefont {N.~E.}\ \bibnamefont {Hussey}},\ }\bibfield  {title}
  {\bibinfo {title} {Quantum oscillations in an overdoped high-${T}_{c}$
  superconductor},\ }\href {https://doi.org/10.1038/nature07323} {\bibfield
  {journal} {\bibinfo  {journal} {Nature}\ }\textbf {\bibinfo {volume} {455}},\
  \bibinfo {pages} {952} (\bibinfo {year} {2008})}\BibitemShut {NoStop}%
\bibitem [{\citenamefont {Bangura}\ \emph {et~al.}(2010)\citenamefont
  {Bangura}, \citenamefont {Rourke}, \citenamefont {Benseman}, \citenamefont
  {Matusiak}, \citenamefont {Cooper}, \citenamefont {Hussey},\ and\
  \citenamefont {Carrington}}]{Bangura2010}%
  \BibitemOpen
  \bibfield  {author} {\bibinfo {author} {\bibfnamefont {A.~F.}\ \bibnamefont
  {Bangura}}, \bibinfo {author} {\bibfnamefont {P.~M.~C.}\ \bibnamefont
  {Rourke}}, \bibinfo {author} {\bibfnamefont {T.~M.}\ \bibnamefont
  {Benseman}}, \bibinfo {author} {\bibfnamefont {M.}~\bibnamefont {Matusiak}},
  \bibinfo {author} {\bibfnamefont {J.~R.}\ \bibnamefont {Cooper}}, \bibinfo
  {author} {\bibfnamefont {N.~E.}\ \bibnamefont {Hussey}},\ and\ \bibinfo
  {author} {\bibfnamefont {A.}~\bibnamefont {Carrington}},\ }\bibfield  {title}
  {\bibinfo {title} {{F}ermi surface and electronic homogeneity of the
  overdoped cuprate superconductor \ce{Tl2Ba2CuO_{6+\delta}} as revealed by
  quantum oscillations},\ }\href {https://doi.org/10.1103/PhysRevB.82.140501}
  {\bibfield  {journal} {\bibinfo  {journal} {Phys. Rev. B}\ }\textbf {\bibinfo
  {volume} {82}},\ \bibinfo {pages} {140501} (\bibinfo {year}
  {2010})}\BibitemShut {NoStop}%
\bibitem [{\citenamefont {Emery}(1987)}]{Emery1987}%
  \BibitemOpen
  \bibfield  {author} {\bibinfo {author} {\bibfnamefont {V.~J.}\ \bibnamefont
  {Emery}},\ }\bibfield  {title} {\bibinfo {title} {Theory of high-${T}_{c}$
  superconductivity in oxides},\ }\href
  {https://doi.org/10.1103/PhysRevLett.58.2794} {\bibfield  {journal} {\bibinfo
   {journal} {Phys. Rev. Lett.}\ }\textbf {\bibinfo {volume} {58}},\ \bibinfo
  {pages} {2794} (\bibinfo {year} {1987})}\BibitemShut {NoStop}%
\bibitem [{\citenamefont {Emery}\ and\ \citenamefont
  {Reiter}(1988)}]{Emery1988}%
  \BibitemOpen
  \bibfield  {author} {\bibinfo {author} {\bibfnamefont {V.~J.}\ \bibnamefont
  {Emery}}\ and\ \bibinfo {author} {\bibfnamefont {G.}~\bibnamefont {Reiter}},\
  }\bibfield  {title} {\bibinfo {title} {Mechanism for high-temperature
  superconductivity},\ }\href {https://doi.org/10.1103/PhysRevB.38.4547}
  {\bibfield  {journal} {\bibinfo  {journal} {Phys. Rev. B}\ }\textbf {\bibinfo
  {volume} {38}},\ \bibinfo {pages} {4547} (\bibinfo {year}
  {1988})}\BibitemShut {NoStop}%
\bibitem [{\citenamefont {Littlewood}\ \emph {et~al.}(1989)\citenamefont
  {Littlewood}, \citenamefont {Varma},\ and\ \citenamefont
  {Abrahams}}]{Littlewood1989}%
  \BibitemOpen
  \bibfield  {author} {\bibinfo {author} {\bibfnamefont {P.~B.}\ \bibnamefont
  {Littlewood}}, \bibinfo {author} {\bibfnamefont {C.~M.}\ \bibnamefont
  {Varma}},\ and\ \bibinfo {author} {\bibfnamefont {E.}~\bibnamefont
  {Abrahams}},\ }\bibfield  {title} {\bibinfo {title} {Pairing instabilities of
  the extended hubbard model for \ce{Cu-O}--based superconductors},\ }\href
  {https://doi.org/10.1103/PhysRevLett.63.2602} {\bibfield  {journal} {\bibinfo
   {journal} {Phys. Rev. Lett.}\ }\textbf {\bibinfo {volume} {63}},\ \bibinfo
  {pages} {2602} (\bibinfo {year} {1989})}\BibitemShut {NoStop}%
\bibitem [{\citenamefont {Gaididei}\ and\ \citenamefont
  {Loktev}(1988)}]{Gaididei1988}%
  \BibitemOpen
  \bibfield  {author} {\bibinfo {author} {\bibfnamefont {Y.~B.}\ \bibnamefont
  {Gaididei}}\ and\ \bibinfo {author} {\bibfnamefont {V.~M.}\ \bibnamefont
  {Loktev}},\ }\bibfield  {title} {\bibinfo {title} {On a theory of the
  electronic spectrum and magnetic properties of high-${T}_{c}$
  superconductors},\ }\href
  {https://doi.org/https://doi.org/10.1002/pssb.2221470135} {\bibfield
  {journal} {\bibinfo  {journal} {Phys. Status Solidi B}\ }\textbf {\bibinfo
  {volume} {147}},\ \bibinfo {pages} {307} (\bibinfo {year}
  {1988})}\BibitemShut {NoStop}%
\bibitem [{\citenamefont {Scalettar}\ \emph {et~al.}(1991)\citenamefont
  {Scalettar}, \citenamefont {Scalapino}, \citenamefont {Sugar},\ and\
  \citenamefont {White}}]{Scalettar1991}%
  \BibitemOpen
  \bibfield  {author} {\bibinfo {author} {\bibfnamefont {R.~T.}\ \bibnamefont
  {Scalettar}}, \bibinfo {author} {\bibfnamefont {D.~J.}\ \bibnamefont
  {Scalapino}}, \bibinfo {author} {\bibfnamefont {R.~L.}\ \bibnamefont
  {Sugar}},\ and\ \bibinfo {author} {\bibfnamefont {S.~R.}\ \bibnamefont
  {White}},\ }\bibfield  {title} {\bibinfo {title} {Antiferromagnetic,
  charge-transfer, and pairing correlations in the three-band {H}ubbard
  model},\ }\href {https://doi.org/10.1103/PhysRevB.44.770} {\bibfield
  {journal} {\bibinfo  {journal} {Phys. Rev. B}\ }\textbf {\bibinfo {volume}
  {44}},\ \bibinfo {pages} {770} (\bibinfo {year} {1991})}\BibitemShut
  {NoStop}%
\bibitem [{\citenamefont {Pickett}(1989)}]{Pickett1989}%
  \BibitemOpen
  \bibfield  {author} {\bibinfo {author} {\bibfnamefont {W.~E.}\ \bibnamefont
  {Pickett}},\ }\bibfield  {title} {\bibinfo {title} {Electronic structure of
  the high-temperature oxide superconductors},\ }\href
  {https://doi.org/10.1103/RevModPhys.61.433} {\bibfield  {journal} {\bibinfo
  {journal} {Rev. Mod. Phys.}\ }\textbf {\bibinfo {volume} {61}},\ \bibinfo
  {pages} {433} (\bibinfo {year} {1989})}\BibitemShut {NoStop}%
\bibitem [{\citenamefont {Pavarini}\ \emph {et~al.}(2001)\citenamefont
  {Pavarini}, \citenamefont {Dasgupta}, \citenamefont {Saha-Dasgupta},
  \citenamefont {Jepsen},\ and\ \citenamefont {Andersen}}]{Pavarini2001}%
  \BibitemOpen
  \bibfield  {author} {\bibinfo {author} {\bibfnamefont {E.}~\bibnamefont
  {Pavarini}}, \bibinfo {author} {\bibfnamefont {I.}~\bibnamefont {Dasgupta}},
  \bibinfo {author} {\bibfnamefont {T.}~\bibnamefont {Saha-Dasgupta}}, \bibinfo
  {author} {\bibfnamefont {O.}~\bibnamefont {Jepsen}},\ and\ \bibinfo {author}
  {\bibfnamefont {O.~K.}\ \bibnamefont {Andersen}},\ }\bibfield  {title}
  {\bibinfo {title} {Band-structure trend in hole-doped cuprates and
  correlation with ${T}_{c~\mathrm{max}}$},\ }\href
  {https://doi.org/10.1103/PhysRevLett.87.047003} {\bibfield  {journal}
  {\bibinfo  {journal} {Phys. Rev. Lett.}\ }\textbf {\bibinfo {volume} {87}},\
  \bibinfo {pages} {047003} (\bibinfo {year} {2001})}\BibitemShut {NoStop}%
\bibitem [{\citenamefont {Kent}\ \emph {et~al.}(2008)\citenamefont {Kent},
  \citenamefont {Saha-Dasgupta}, \citenamefont {Jepsen}, \citenamefont
  {Andersen}, \citenamefont {Macridin}, \citenamefont {Maier}, \citenamefont
  {Jarrell},\ and\ \citenamefont {Schulthess}}]{Kent2008}%
  \BibitemOpen
  \bibfield  {author} {\bibinfo {author} {\bibfnamefont {P.~R.~C.}\
  \bibnamefont {Kent}}, \bibinfo {author} {\bibfnamefont {T.}~\bibnamefont
  {Saha-Dasgupta}}, \bibinfo {author} {\bibfnamefont {O.}~\bibnamefont
  {Jepsen}}, \bibinfo {author} {\bibfnamefont {O.~K.}\ \bibnamefont
  {Andersen}}, \bibinfo {author} {\bibfnamefont {A.}~\bibnamefont {Macridin}},
  \bibinfo {author} {\bibfnamefont {T.~A.}\ \bibnamefont {Maier}}, \bibinfo
  {author} {\bibfnamefont {M.}~\bibnamefont {Jarrell}},\ and\ \bibinfo {author}
  {\bibfnamefont {T.~C.}\ \bibnamefont {Schulthess}},\ }\bibfield  {title}
  {\bibinfo {title} {Combined density functional and dynamical cluster quantum
  {M}onte {C}arlo calculations of the three-band {H}ubbard model for hole-doped
  cuprate superconductors},\ }\href
  {https://doi.org/10.1103/PhysRevB.78.035132} {\bibfield  {journal} {\bibinfo
  {journal} {Phys. Rev. B}\ }\textbf {\bibinfo {volume} {78}},\ \bibinfo
  {pages} {035132} (\bibinfo {year} {2008})}\BibitemShut {NoStop}%
\bibitem [{\citenamefont {Weber}\ \emph {et~al.}(2014)\citenamefont {Weber},
  \citenamefont {Giamarchi},\ and\ \citenamefont {Varma}}]{Weber2014}%
  \BibitemOpen
  \bibfield  {author} {\bibinfo {author} {\bibfnamefont {C.}~\bibnamefont
  {Weber}}, \bibinfo {author} {\bibfnamefont {T.}~\bibnamefont {Giamarchi}},\
  and\ \bibinfo {author} {\bibfnamefont {C.~M.}\ \bibnamefont {Varma}},\
  }\bibfield  {title} {\bibinfo {title} {Phase diagram of a three-orbital model
  for high-${T}_{c}$ cuprate superconductors},\ }\href
  {https://doi.org/10.1103/PhysRevLett.112.117001} {\bibfield  {journal}
  {\bibinfo  {journal} {Phys. Rev. Lett.}\ }\textbf {\bibinfo {volume} {112}},\
  \bibinfo {pages} {117001} (\bibinfo {year} {2014})}\BibitemShut {NoStop}%
\bibitem [{\citenamefont {Photopoulos}\ and\ \citenamefont
  {Fr\'{e}sard}(2019)}]{Photopoulos2019}%
  \BibitemOpen
  \bibfield  {author} {\bibinfo {author} {\bibfnamefont {R.}~\bibnamefont
  {Photopoulos}}\ and\ \bibinfo {author} {\bibfnamefont {R.}~\bibnamefont
  {Fr\'{e}sard}},\ }\bibfield  {title} {\bibinfo {title} {A {3D} tight-binding
  model for \ce{La}-based cuprate superconductors},\ }\href
  {https://doi.org/https://doi.org/10.1002/andp.201900177} {\bibfield
  {journal} {\bibinfo  {journal} {Ann. Phys. (Berl.)}\ }\textbf {\bibinfo
  {volume} {531}},\ \bibinfo {pages} {1900177} (\bibinfo {year}
  {2019})}\BibitemShut {NoStop}%
\bibitem [{\citenamefont {Weber}\ \emph {et~al.}(2009)\citenamefont {Weber},
  \citenamefont {L\"auchli}, \citenamefont {Mila},\ and\ \citenamefont
  {Giamarchi}}]{Weber2009}%
  \BibitemOpen
  \bibfield  {author} {\bibinfo {author} {\bibfnamefont {C.}~\bibnamefont
  {Weber}}, \bibinfo {author} {\bibfnamefont {A.}~\bibnamefont {L\"auchli}},
  \bibinfo {author} {\bibfnamefont {F.}~\bibnamefont {Mila}},\ and\ \bibinfo
  {author} {\bibfnamefont {T.}~\bibnamefont {Giamarchi}},\ }\bibfield  {title}
  {\bibinfo {title} {Orbital currents in extended {H}ubbard models of
  high-${T}_{c}$ cuprate superconductors},\ }\href
  {https://doi.org/10.1103/PhysRevLett.102.017005} {\bibfield  {journal}
  {\bibinfo  {journal} {Phys. Rev. Lett.}\ }\textbf {\bibinfo {volume} {102}},\
  \bibinfo {pages} {017005} (\bibinfo {year} {2009})}\BibitemShut {NoStop}%
\bibitem [{\citenamefont {Tazai}\ \emph {et~al.}(2021)\citenamefont {Tazai},
  \citenamefont {Yamakawa},\ and\ \citenamefont {Kontani}}]{Tazai2021}%
  \BibitemOpen
  \bibfield  {author} {\bibinfo {author} {\bibfnamefont {R.}~\bibnamefont
  {Tazai}}, \bibinfo {author} {\bibfnamefont {Y.}~\bibnamefont {Yamakawa}},\
  and\ \bibinfo {author} {\bibfnamefont {H.}~\bibnamefont {Kontani}},\
  }\bibfield  {title} {\bibinfo {title} {Emergence of charge loop current in
  the geometrically frustrated hubbard model: A functional renormalization
  group study},\ }\href {https://doi.org/10.1103/PhysRevB.103.L161112}
  {\bibfield  {journal} {\bibinfo  {journal} {Phys. Rev. B}\ }\textbf {\bibinfo
  {volume} {103}},\ \bibinfo {pages} {L161112} (\bibinfo {year}
  {2021})}\BibitemShut {NoStop}%
\bibitem [{\citenamefont {Greiter}\ and\ \citenamefont
  {Thomale}(2007)}]{Greiter2007}%
  \BibitemOpen
  \bibfield  {author} {\bibinfo {author} {\bibfnamefont {M.}~\bibnamefont
  {Greiter}}\ and\ \bibinfo {author} {\bibfnamefont {R.}~\bibnamefont
  {Thomale}},\ }\bibfield  {title} {\bibinfo {title} {No evidence for
  spontaneous orbital currents in numerical studies of three-band models for
  the \ce{CuO} planes of high temperature superconductors},\ }\href
  {https://doi.org/10.1103/PhysRevLett.99.027005} {\bibfield  {journal}
  {\bibinfo  {journal} {Phys. Rev. Lett.}\ }\textbf {\bibinfo {volume} {99}},\
  \bibinfo {pages} {027005} (\bibinfo {year} {2007})}\BibitemShut {NoStop}%
\bibitem [{\citenamefont {Thomale}\ and\ \citenamefont
  {Greiter}(2008)}]{Thomale2008}%
  \BibitemOpen
  \bibfield  {author} {\bibinfo {author} {\bibfnamefont {R.}~\bibnamefont
  {Thomale}}\ and\ \bibinfo {author} {\bibfnamefont {M.}~\bibnamefont
  {Greiter}},\ }\bibfield  {title} {\bibinfo {title} {Numerical analysis of
  three-band models for \ce{CuO} planes as candidates for a spontaneous
  {T}-violating orbital current phase},\ }\href
  {https://doi.org/10.1103/PhysRevB.77.094511} {\bibfield  {journal} {\bibinfo
  {journal} {Phys. Rev. B}\ }\textbf {\bibinfo {volume} {77}},\ \bibinfo
  {pages} {094511} (\bibinfo {year} {2008})}\BibitemShut {NoStop}%
\bibitem [{\citenamefont {Kung}\ \emph {et~al.}(2014)\citenamefont {Kung},
  \citenamefont {Chen}, \citenamefont {Moritz}, \citenamefont {Johnston},
  \citenamefont {Thomale},\ and\ \citenamefont {Devereaux}}]{Kung2014}%
  \BibitemOpen
  \bibfield  {author} {\bibinfo {author} {\bibfnamefont {Y.~F.}\ \bibnamefont
  {Kung}}, \bibinfo {author} {\bibfnamefont {C.-C.}\ \bibnamefont {Chen}},
  \bibinfo {author} {\bibfnamefont {B.}~\bibnamefont {Moritz}}, \bibinfo
  {author} {\bibfnamefont {S.}~\bibnamefont {Johnston}}, \bibinfo {author}
  {\bibfnamefont {R.}~\bibnamefont {Thomale}},\ and\ \bibinfo {author}
  {\bibfnamefont {T.~P.}\ \bibnamefont {Devereaux}},\ }\bibfield  {title}
  {\bibinfo {title} {Numerical exploration of spontaneous broken symmetries in
  multiorbital {H}ubbard models},\ }\href
  {https://doi.org/10.1103/PhysRevB.90.224507} {\bibfield  {journal} {\bibinfo
  {journal} {Phys. Rev. B}\ }\textbf {\bibinfo {volume} {90}},\ \bibinfo
  {pages} {224507} (\bibinfo {year} {2014})}\BibitemShut {NoStop}%
\bibitem [{\citenamefont {Dresselhaus}\ \emph {et~al.}(2008)\citenamefont
  {Dresselhaus}, \citenamefont {Dresselhaus},\ and\ \citenamefont
  {Jorio}}]{Dresselhaus2007}%
  \BibitemOpen
  \bibfield  {author} {\bibinfo {author} {\bibfnamefont {M.~S.}\ \bibnamefont
  {Dresselhaus}}, \bibinfo {author} {\bibfnamefont {G.}~\bibnamefont
  {Dresselhaus}},\ and\ \bibinfo {author} {\bibfnamefont {A.}~\bibnamefont
  {Jorio}},\ }\href {https://doi.org/10.1007/978-3-540-32899-5} {\emph
  {\bibinfo {title} {Group theory: Application to the Physics of Condensed
  Matter}}}\ (\bibinfo  {publisher} {Springer},\ \bibinfo {address} {Berlin
  Heidelberg},\ \bibinfo {year} {2008})\BibitemShut {NoStop}%
\bibitem [{\citenamefont {Andersen}\ \emph {et~al.}(1995)\citenamefont
  {Andersen}, \citenamefont {Liechtenstein}, \citenamefont {Jepsen},\ and\
  \citenamefont {Paulsen}}]{Andersen1995}%
  \BibitemOpen
  \bibfield  {author} {\bibinfo {author} {\bibfnamefont {O.}~\bibnamefont
  {Andersen}}, \bibinfo {author} {\bibfnamefont {A.}~\bibnamefont
  {Liechtenstein}}, \bibinfo {author} {\bibfnamefont {O.}~\bibnamefont
  {Jepsen}},\ and\ \bibinfo {author} {\bibfnamefont {F.}~\bibnamefont
  {Paulsen}},\ }\bibfield  {title} {\bibinfo {title} {{LDA} energy bands,
  low-energy {H}amiltonians, ${t'}$, ${t''}$, ${t_{\perp}(\mathbf{k})}$, and
  ${J_{\perp}}$},\ }\href
  {https://doi.org/https://doi.org/10.1016/0022-3697(95)00269-3} {\bibfield
  {journal} {\bibinfo  {journal} {J. Phys. Chem. Solids}\ }\textbf {\bibinfo
  {volume} {56}},\ \bibinfo {pages} {1573} (\bibinfo {year} {1995})},\ \bibinfo
  {note} {proceedings of the Conference on Spectroscopies in Novel
  Superconductors}\BibitemShut {NoStop}%
\bibitem [{\citenamefont {Klug}\ \emph {et~al.}(2018)\citenamefont {Klug},
  \citenamefont {Kang}, \citenamefont {Fernandes},\ and\ \citenamefont
  {Schmalian}}]{Klug2018}%
  \BibitemOpen
  \bibfield  {author} {\bibinfo {author} {\bibfnamefont {M.}~\bibnamefont
  {Klug}}, \bibinfo {author} {\bibfnamefont {J.}~\bibnamefont {Kang}}, \bibinfo
  {author} {\bibfnamefont {R.~M.}\ \bibnamefont {Fernandes}},\ and\ \bibinfo
  {author} {\bibfnamefont {J.}~\bibnamefont {Schmalian}},\ }\bibfield  {title}
  {\bibinfo {title} {Orbital loop currents in iron-based superconductors},\
  }\href {https://doi.org/10.1103/PhysRevB.97.155130} {\bibfield  {journal}
  {\bibinfo  {journal} {Phys. Rev. B}\ }\textbf {\bibinfo {volume} {97}},\
  \bibinfo {pages} {155130} (\bibinfo {year} {2018})}\BibitemShut {NoStop}%
\bibitem [{\citenamefont {Christensen}\ \emph {et~al.}(2022)\citenamefont
  {Christensen}, \citenamefont {Birol}, \citenamefont {Andersen},\ and\
  \citenamefont {Fernandes}}]{Christensen2022}%
  \BibitemOpen
  \bibfield  {author} {\bibinfo {author} {\bibfnamefont {M.~H.}\ \bibnamefont
  {Christensen}}, \bibinfo {author} {\bibfnamefont {T.}~\bibnamefont {Birol}},
  \bibinfo {author} {\bibfnamefont {B.~M.}\ \bibnamefont {Andersen}},\ and\
  \bibinfo {author} {\bibfnamefont {R.~M.}\ \bibnamefont {Fernandes}},\
  }\bibfield  {title} {\bibinfo {title} {Loop currents in \ce{AV3Sb5} kagome
  metals: Multipolar and toroidal magnetic orders},\ }\href
  {https://doi.org/10.1103/PhysRevB.106.144504} {\bibfield  {journal} {\bibinfo
   {journal} {Phys. Rev. B}\ }\textbf {\bibinfo {volume} {106}},\ \bibinfo
  {pages} {144504} (\bibinfo {year} {2022})}\BibitemShut {NoStop}%
\bibitem [{\citenamefont {Kramer}\ \emph {et~al.}(2019)\citenamefont {Kramer},
  \citenamefont {Horio}, \citenamefont {Tsirkin}, \citenamefont {Sassa},
  \citenamefont {Hauser}, \citenamefont {Matt}, \citenamefont {Sutter},
  \citenamefont {Chikina}, \citenamefont {Schr\"{o}ter}, \citenamefont
  {Krieger}, \citenamefont {Schmitt}, \citenamefont {Strocov}, \citenamefont
  {Plumb}, \citenamefont {Shi}, \citenamefont {Pyon}, \citenamefont {Takayama},
  \citenamefont {Takagi}, \citenamefont {Adachi}, \citenamefont {Ohgi},
  \citenamefont {Kawamata}, \citenamefont {Koike}, \citenamefont {Kondo},
  \citenamefont {Lipscombe}, \citenamefont {Hayden}, \citenamefont {Ishikado},
  \citenamefont {Eisaki}, \citenamefont {Neupert},\ and\ \citenamefont
  {Chang}}]{Kramer2019}%
  \BibitemOpen
  \bibfield  {author} {\bibinfo {author} {\bibfnamefont {K.~P.}\ \bibnamefont
  {Kramer}}, \bibinfo {author} {\bibfnamefont {M.}~\bibnamefont {Horio}},
  \bibinfo {author} {\bibfnamefont {S.~S.}\ \bibnamefont {Tsirkin}}, \bibinfo
  {author} {\bibfnamefont {Y.}~\bibnamefont {Sassa}}, \bibinfo {author}
  {\bibfnamefont {K.}~\bibnamefont {Hauser}}, \bibinfo {author} {\bibfnamefont
  {C.~E.}\ \bibnamefont {Matt}}, \bibinfo {author} {\bibfnamefont
  {D.}~\bibnamefont {Sutter}}, \bibinfo {author} {\bibfnamefont
  {A.}~\bibnamefont {Chikina}}, \bibinfo {author} {\bibfnamefont {N.~B.~M.}\
  \bibnamefont {Schr\"{o}ter}}, \bibinfo {author} {\bibfnamefont {J.~A.}\
  \bibnamefont {Krieger}}, \bibinfo {author} {\bibfnamefont {T.}~\bibnamefont
  {Schmitt}}, \bibinfo {author} {\bibfnamefont {V.~N.}\ \bibnamefont
  {Strocov}}, \bibinfo {author} {\bibfnamefont {N.~C.}\ \bibnamefont {Plumb}},
  \bibinfo {author} {\bibfnamefont {M.}~\bibnamefont {Shi}}, \bibinfo {author}
  {\bibfnamefont {S.}~\bibnamefont {Pyon}}, \bibinfo {author} {\bibfnamefont
  {T.}~\bibnamefont {Takayama}}, \bibinfo {author} {\bibfnamefont
  {H.}~\bibnamefont {Takagi}}, \bibinfo {author} {\bibfnamefont
  {T.}~\bibnamefont {Adachi}}, \bibinfo {author} {\bibfnamefont
  {T.}~\bibnamefont {Ohgi}}, \bibinfo {author} {\bibfnamefont {T.}~\bibnamefont
  {Kawamata}}, \bibinfo {author} {\bibfnamefont {Y.}~\bibnamefont {Koike}},
  \bibinfo {author} {\bibfnamefont {T.}~\bibnamefont {Kondo}}, \bibinfo
  {author} {\bibfnamefont {O.~J.}\ \bibnamefont {Lipscombe}}, \bibinfo {author}
  {\bibfnamefont {S.~M.}\ \bibnamefont {Hayden}}, \bibinfo {author}
  {\bibfnamefont {M.}~\bibnamefont {Ishikado}}, \bibinfo {author}
  {\bibfnamefont {H.}~\bibnamefont {Eisaki}}, \bibinfo {author} {\bibfnamefont
  {T.}~\bibnamefont {Neupert}},\ and\ \bibinfo {author} {\bibfnamefont
  {J.}~\bibnamefont {Chang}},\ }\bibfield  {title} {\bibinfo {title} {Band
  structure of overdoped cuprate superconductors: Density functional theory
  matching experiments},\ }\href {https://doi.org/10.1103/PhysRevB.99.224509}
  {\bibfield  {journal} {\bibinfo  {journal} {Phys. Rev. B}\ }\textbf {\bibinfo
  {volume} {99}},\ \bibinfo {pages} {224509} (\bibinfo {year}
  {2019})}\BibitemShut {NoStop}%
\bibitem [{\citenamefont {Lee-Hone}\ \emph {et~al.}(2020)\citenamefont
  {Lee-Hone}, \citenamefont {\"{O}zdemir}, \citenamefont {Mishra},
  \citenamefont {Broun},\ and\ \citenamefont {Hirschfeld}}]{Lee-Hone2020}%
  \BibitemOpen
  \bibfield  {author} {\bibinfo {author} {\bibfnamefont {N.~R.}\ \bibnamefont
  {Lee-Hone}}, \bibinfo {author} {\bibfnamefont {H.~U.}\ \bibnamefont
  {\"{O}zdemir}}, \bibinfo {author} {\bibfnamefont {V.}~\bibnamefont {Mishra}},
  \bibinfo {author} {\bibfnamefont {D.~M.}\ \bibnamefont {Broun}},\ and\
  \bibinfo {author} {\bibfnamefont {P.~J.}\ \bibnamefont {Hirschfeld}},\
  }\bibfield  {title} {\bibinfo {title} {Low energy phenomenology of the
  overdoped cuprates: Viability of the {L}andau-{BCS} paradigm},\ }\href
  {https://doi.org/10.1103/PhysRevResearch.2.013228} {\bibfield  {journal}
  {\bibinfo  {journal} {Phys. Rev. Res.}\ }\textbf {\bibinfo {volume} {2}},\
  \bibinfo {pages} {013228} (\bibinfo {year} {2020})}\BibitemShut {NoStop}%
\bibitem [{\citenamefont {Chakravarty}\ \emph
  {et~al.}(2001{\natexlab{a}})\citenamefont {Chakravarty}, \citenamefont
  {Laughlin}, \citenamefont {Morr},\ and\ \citenamefont {Nayak}}]{Morr2001}%
  \BibitemOpen
  \bibfield  {author} {\bibinfo {author} {\bibfnamefont {S.}~\bibnamefont
  {Chakravarty}}, \bibinfo {author} {\bibfnamefont {R.~B.}\ \bibnamefont
  {Laughlin}}, \bibinfo {author} {\bibfnamefont {D.~K.}\ \bibnamefont {Morr}},\
  and\ \bibinfo {author} {\bibfnamefont {C.}~\bibnamefont {Nayak}},\ }\bibfield
   {title} {\bibinfo {title} {Hidden order in the cuprates},\ }\href
  {https://doi.org/10.1103/PhysRevB.63.094503} {\bibfield  {journal} {\bibinfo
  {journal} {Phys. Rev. B}\ }\textbf {\bibinfo {volume} {63}},\ \bibinfo
  {pages} {094503} (\bibinfo {year} {2001}{\natexlab{a}})}\BibitemShut
  {NoStop}%
\bibitem [{\citenamefont {Chakravarty}\ \emph
  {et~al.}(2001{\natexlab{b}})\citenamefont {Chakravarty}, \citenamefont
  {Laughlin}, \citenamefont {Morr},\ and\ \citenamefont
  {Nayak}}]{Chakravarty2001}%
  \BibitemOpen
  \bibfield  {author} {\bibinfo {author} {\bibfnamefont {S.}~\bibnamefont
  {Chakravarty}}, \bibinfo {author} {\bibfnamefont {R.~B.}\ \bibnamefont
  {Laughlin}}, \bibinfo {author} {\bibfnamefont {D.~K.}\ \bibnamefont {Morr}},\
  and\ \bibinfo {author} {\bibfnamefont {C.}~\bibnamefont {Nayak}},\ }\bibfield
   {title} {\bibinfo {title} {Hidden order in the cuprates},\ }\href
  {https://doi.org/10.1103/PhysRevB.63.094503} {\bibfield  {journal} {\bibinfo
  {journal} {Phys. Rev. B}\ }\textbf {\bibinfo {volume} {63}},\ \bibinfo
  {pages} {094503} (\bibinfo {year} {2001}{\natexlab{b}})}\BibitemShut
  {NoStop}%
\bibitem [{\citenamefont {Varma}\ and\ \citenamefont {Zhu}(2015)}]{Varma2015}%
  \BibitemOpen
  \bibfield  {author} {\bibinfo {author} {\bibfnamefont {C.~M.}\ \bibnamefont
  {Varma}}\ and\ \bibinfo {author} {\bibfnamefont {L.}~\bibnamefont {Zhu}},\
  }\bibfield  {title} {\bibinfo {title} {Specific heat and sound velocity at
  the relevant competing phase of high-temperature superconductors},\ }\href
  {https://doi.org/10.1073/pnas.1417150112} {\bibfield  {journal} {\bibinfo
  {journal} {PNAS}\ }\textbf {\bibinfo {volume} {112}},\ \bibinfo {pages}
  {6331} (\bibinfo {year} {2015})}\BibitemShut {NoStop}%
\bibitem [{\citenamefont {Varma}(2019)}]{Varma2019}%
  \BibitemOpen
  \bibfield  {author} {\bibinfo {author} {\bibfnamefont {C.~M.}\ \bibnamefont
  {Varma}},\ }\bibfield  {title} {\bibinfo {title} {Pseudogap and {F}ermi arcs
  in underdoped cuprates},\ }\href {https://doi.org/10.1103/PhysRevB.99.224516}
  {\bibfield  {journal} {\bibinfo  {journal} {Phys. Rev. B}\ }\textbf {\bibinfo
  {volume} {99}},\ \bibinfo {pages} {224516} (\bibinfo {year}
  {2019})}\BibitemShut {NoStop}%
\bibitem [{\citenamefont {Bounoua}\ \emph {et~al.}(2022)\citenamefont
  {Bounoua}, \citenamefont {Sidis}, \citenamefont {Loew}, \citenamefont
  {Bourdarot}, \citenamefont {Boehm}, \citenamefont {Steffens}, \citenamefont
  {Mangin-Thro}, \citenamefont {Bal{\'e}dent},\ and\ \citenamefont
  {Bourges}}]{Bounoua2022}%
  \BibitemOpen
  \bibfield  {author} {\bibinfo {author} {\bibfnamefont {D.}~\bibnamefont
  {Bounoua}}, \bibinfo {author} {\bibfnamefont {Y.}~\bibnamefont {Sidis}},
  \bibinfo {author} {\bibfnamefont {T.}~\bibnamefont {Loew}}, \bibinfo {author}
  {\bibfnamefont {F.}~\bibnamefont {Bourdarot}}, \bibinfo {author}
  {\bibfnamefont {M.}~\bibnamefont {Boehm}}, \bibinfo {author} {\bibfnamefont
  {P.}~\bibnamefont {Steffens}}, \bibinfo {author} {\bibfnamefont
  {L.}~\bibnamefont {Mangin-Thro}}, \bibinfo {author} {\bibfnamefont
  {V.}~\bibnamefont {Bal{\'e}dent}},\ and\ \bibinfo {author} {\bibfnamefont
  {P.}~\bibnamefont {Bourges}},\ }\bibfield  {title} {\bibinfo {title} {Hidden
  magnetic texture in the pseudogap phase of high-${T}_{c}$
  \ce{YBa2Cu3O_{6.6}}},\ }\href {https://doi.org/10.1038/s42005-022-01048-1}
  {\bibfield  {journal} {\bibinfo  {journal} {Commun. Phys.}\ }\textbf
  {\bibinfo {volume} {5}},\ \bibinfo {pages} {268} (\bibinfo {year}
  {2022})}\BibitemShut {NoStop}%
\bibitem [{\citenamefont {Bounoua}\ \emph {et~al.}(2023)\citenamefont
  {Bounoua}, \citenamefont {Sidis}, \citenamefont {Boehm}, \citenamefont
  {Steffens}, \citenamefont {Loew}, \citenamefont {Guo}, \citenamefont {Qian},
  \citenamefont {Yao},\ and\ \citenamefont {Bourges}}]{Bounoua2023}%
  \BibitemOpen
  \bibfield  {author} {\bibinfo {author} {\bibfnamefont {D.}~\bibnamefont
  {Bounoua}}, \bibinfo {author} {\bibfnamefont {Y.}~\bibnamefont {Sidis}},
  \bibinfo {author} {\bibfnamefont {M.}~\bibnamefont {Boehm}}, \bibinfo
  {author} {\bibfnamefont {P.}~\bibnamefont {Steffens}}, \bibinfo {author}
  {\bibfnamefont {T.}~\bibnamefont {Loew}}, \bibinfo {author} {\bibfnamefont
  {L.~S.}\ \bibnamefont {Guo}}, \bibinfo {author} {\bibfnamefont
  {J.}~\bibnamefont {Qian}}, \bibinfo {author} {\bibfnamefont {X.}~\bibnamefont
  {Yao}},\ and\ \bibinfo {author} {\bibfnamefont {P.}~\bibnamefont {Bourges}},\
  }\bibfield  {title} {\bibinfo {title} {Universality of the $\mathbf{q}=1/2$
  orbital magnetism in the pseudogap phase of the high-${T}_{c}$ superconductor
  \ce{YBa2Cu3O_{6+x}}},\ }\href@noop {} {\bibfield  {journal} {\bibinfo
  {journal} {\href{https://arxiv.org/abs/2302.01870}{arXiv:2302.01870}
  [cond-mat.str-el]}\ } (\bibinfo {year} {2023})}\BibitemShut {NoStop}%
\bibitem [{\citenamefont {Scalapino}(1995)}]{scalapino1995case}%
  \BibitemOpen
  \bibfield  {author} {\bibinfo {author} {\bibfnamefont {D.~J.}\ \bibnamefont
  {Scalapino}},\ }\bibfield  {title} {\bibinfo {title} {The case for
  $d_{x^2-y^2}$ pairing in the cuprate superconductors},\ }\href
  {https://doi.org/10.1016/0370-1573(94)00086-I} {\bibfield  {journal}
  {\bibinfo  {journal} {Phys. Rep.}\ }\textbf {\bibinfo {volume} {250}},\
  \bibinfo {pages} {329} (\bibinfo {year} {1995})}\BibitemShut {NoStop}%
\bibitem [{\citenamefont {Maiti}\ and\ \citenamefont
  {Chubukov}(2013)}]{Maiti2013}%
  \BibitemOpen
  \bibfield  {author} {\bibinfo {author} {\bibfnamefont {S.}~\bibnamefont
  {Maiti}}\ and\ \bibinfo {author} {\bibfnamefont {A.~V.}\ \bibnamefont
  {Chubukov}},\ }\bibfield  {title} {\bibinfo {title} {{Superconductivity from
  repulsive interaction}},\ }\href {https://doi.org/10.1063/1.4818400}
  {\bibfield  {journal} {\bibinfo  {journal} {AIP Conf. Proc.}\ }\textbf
  {\bibinfo {volume} {1550}},\ \bibinfo {pages} {3} (\bibinfo {year}
  {2013})}\BibitemShut {NoStop}%
\bibitem [{\citenamefont {Leggett}(2006)}]{Leggett2006}%
  \BibitemOpen
  \bibfield  {author} {\bibinfo {author} {\bibfnamefont {A.~J.}\ \bibnamefont
  {Leggett}},\ }\href
  {https://doi.org/10.1093/acprof:oso/9780198526438.001.0001} {\emph {\bibinfo
  {title} {{Quantum Liquids: {B}ose condensation and {C}ooper pairing in
  condensed-matter systems}}}}\ (\bibinfo  {publisher} {Oxford University
  Press},\ \bibinfo {address} {Oxford},\ \bibinfo {year} {2006})\BibitemShut
  {NoStop}%
\bibitem [{\citenamefont {Sigrist}(2005)}]{Sigrist2005}%
  \BibitemOpen
  \bibfield  {author} {\bibinfo {author} {\bibfnamefont {M.}~\bibnamefont
  {Sigrist}},\ }\bibfield  {title} {\bibinfo {title} {{Introduction to
  Unconventional Superconductivity}},\ }\href
  {https://doi.org/10.1063/1.2080350} {\bibfield  {journal} {\bibinfo
  {journal} {AIP Conf. Proc.}\ }\textbf {\bibinfo {volume} {789}},\ \bibinfo
  {pages} {165} (\bibinfo {year} {2005})}\BibitemShut {NoStop}%
\bibitem [{\citenamefont {Dagotto}(1994)}]{Dagotto1994}%
  \BibitemOpen
  \bibfield  {author} {\bibinfo {author} {\bibfnamefont {E.}~\bibnamefont
  {Dagotto}},\ }\bibfield  {title} {\bibinfo {title} {Correlated electrons in
  high-temperature superconductors},\ }\href
  {https://doi.org/10.1103/RevModPhys.66.763} {\bibfield  {journal} {\bibinfo
  {journal} {Rev. Mod. Phys.}\ }\textbf {\bibinfo {volume} {66}},\ \bibinfo
  {pages} {763} (\bibinfo {year} {1994})}\BibitemShut {NoStop}%
\bibitem [{\citenamefont {Pellegrin}\ \emph {et~al.}(1993)\citenamefont
  {Pellegrin}, \citenamefont {N\"{u}cker}, \citenamefont {Fink}, \citenamefont
  {Molodtsov}, \citenamefont {Guti\'{e}rrez}, \citenamefont {Navas},
  \citenamefont {Strebel}, \citenamefont {Hu}, \citenamefont {Domke},
  \citenamefont {Kaindl}, \citenamefont {Uchida}, \citenamefont {Nakamura},
  \citenamefont {Markl}, \citenamefont {Klauda}, \citenamefont
  {Saemann-Ischenko}, \citenamefont {Krol}, \citenamefont {Peng}, \citenamefont
  {Li},\ and\ \citenamefont {Greene}}]{Pellegrin1993}%
  \BibitemOpen
  \bibfield  {author} {\bibinfo {author} {\bibfnamefont {E.}~\bibnamefont
  {Pellegrin}}, \bibinfo {author} {\bibfnamefont {N.}~\bibnamefont
  {N\"{u}cker}}, \bibinfo {author} {\bibfnamefont {J.}~\bibnamefont {Fink}},
  \bibinfo {author} {\bibfnamefont {S.~L.}\ \bibnamefont {Molodtsov}}, \bibinfo
  {author} {\bibfnamefont {A.}~\bibnamefont {Guti\'{e}rrez}}, \bibinfo {author}
  {\bibfnamefont {E.}~\bibnamefont {Navas}}, \bibinfo {author} {\bibfnamefont
  {O.}~\bibnamefont {Strebel}}, \bibinfo {author} {\bibfnamefont
  {Z.}~\bibnamefont {Hu}}, \bibinfo {author} {\bibfnamefont {M.}~\bibnamefont
  {Domke}}, \bibinfo {author} {\bibfnamefont {G.}~\bibnamefont {Kaindl}},
  \bibinfo {author} {\bibfnamefont {S.}~\bibnamefont {Uchida}}, \bibinfo
  {author} {\bibfnamefont {Y.}~\bibnamefont {Nakamura}}, \bibinfo {author}
  {\bibfnamefont {J.}~\bibnamefont {Markl}}, \bibinfo {author} {\bibfnamefont
  {M.}~\bibnamefont {Klauda}}, \bibinfo {author} {\bibfnamefont
  {G.}~\bibnamefont {Saemann-Ischenko}}, \bibinfo {author} {\bibfnamefont
  {A.}~\bibnamefont {Krol}}, \bibinfo {author} {\bibfnamefont {J.~L.}\
  \bibnamefont {Peng}}, \bibinfo {author} {\bibfnamefont {Z.~Y.}\ \bibnamefont
  {Li}},\ and\ \bibinfo {author} {\bibfnamefont {R.~L.}\ \bibnamefont
  {Greene}},\ }\bibfield  {title} {\bibinfo {title} {Orbital character of
  states at the {F}ermi level in \ce{La_{2-x}Sr_xCuO4} and \ce{R_{2-x}Ce_xCuO4}
  (\ce{R}=\ce{Nd},\ce{Sm})},\ }\href {https://doi.org/10.1103/PhysRevB.47.3354}
  {\bibfield  {journal} {\bibinfo  {journal} {Phys. Rev. B}\ }\textbf {\bibinfo
  {volume} {47}},\ \bibinfo {pages} {3354} (\bibinfo {year}
  {1993})}\BibitemShut {NoStop}%
\bibitem [{\citenamefont {Yoshida}\ \emph {et~al.}(2006)\citenamefont
  {Yoshida}, \citenamefont {Zhou}, \citenamefont {Tanaka}, \citenamefont
  {Yang}, \citenamefont {Hussain}, \citenamefont {Shen}, \citenamefont
  {Fujimori}, \citenamefont {Sahrakorpi}, \citenamefont {Lindroos},
  \citenamefont {Markiewicz}, \citenamefont {Bansil}, \citenamefont {Komiya},
  \citenamefont {Ando}, \citenamefont {Eisaki}, \citenamefont {Kakeshita},\
  and\ \citenamefont {Uchida}}]{Yoshida2006}%
  \BibitemOpen
  \bibfield  {author} {\bibinfo {author} {\bibfnamefont {T.}~\bibnamefont
  {Yoshida}}, \bibinfo {author} {\bibfnamefont {X.~J.}\ \bibnamefont {Zhou}},
  \bibinfo {author} {\bibfnamefont {K.}~\bibnamefont {Tanaka}}, \bibinfo
  {author} {\bibfnamefont {W.~L.}\ \bibnamefont {Yang}}, \bibinfo {author}
  {\bibfnamefont {Z.}~\bibnamefont {Hussain}}, \bibinfo {author} {\bibfnamefont
  {Z.-X.}\ \bibnamefont {Shen}}, \bibinfo {author} {\bibfnamefont
  {A.}~\bibnamefont {Fujimori}}, \bibinfo {author} {\bibfnamefont
  {S.}~\bibnamefont {Sahrakorpi}}, \bibinfo {author} {\bibfnamefont
  {M.}~\bibnamefont {Lindroos}}, \bibinfo {author} {\bibfnamefont {R.~S.}\
  \bibnamefont {Markiewicz}}, \bibinfo {author} {\bibfnamefont
  {A.}~\bibnamefont {Bansil}}, \bibinfo {author} {\bibfnamefont
  {S.}~\bibnamefont {Komiya}}, \bibinfo {author} {\bibfnamefont
  {Y.}~\bibnamefont {Ando}}, \bibinfo {author} {\bibfnamefont {H.}~\bibnamefont
  {Eisaki}}, \bibinfo {author} {\bibfnamefont {T.}~\bibnamefont {Kakeshita}},\
  and\ \bibinfo {author} {\bibfnamefont {S.}~\bibnamefont {Uchida}},\
  }\bibfield  {title} {\bibinfo {title} {Systematic doping evolution of the
  underlying {F}ermi surface of \ce{La_{2-x}Sr_{x}CuO4}},\ }\href
  {https://doi.org/10.1103/PhysRevB.74.224510} {\bibfield  {journal} {\bibinfo
  {journal} {Phys. Rev. B}\ }\textbf {\bibinfo {volume} {74}},\ \bibinfo
  {pages} {224510} (\bibinfo {year} {2006})}\BibitemShut {NoStop}%
\bibitem [{\citenamefont {Peets}\ \emph {et~al.}(2007)\citenamefont {Peets},
  \citenamefont {Mottershead}, \citenamefont {Wu}, \citenamefont {Elfimov},
  \citenamefont {Liang}, \citenamefont {Hardy}, \citenamefont {Bonn},
  \citenamefont {Raudsepp}, \citenamefont {Ingle},\ and\ \citenamefont
  {Damascelli}}]{Peets2007}%
  \BibitemOpen
  \bibfield  {author} {\bibinfo {author} {\bibfnamefont {D.~C.}\ \bibnamefont
  {Peets}}, \bibinfo {author} {\bibfnamefont {J.~D.~F.}\ \bibnamefont
  {Mottershead}}, \bibinfo {author} {\bibfnamefont {B.}~\bibnamefont {Wu}},
  \bibinfo {author} {\bibfnamefont {I.~S.}\ \bibnamefont {Elfimov}}, \bibinfo
  {author} {\bibfnamefont {R.}~\bibnamefont {Liang}}, \bibinfo {author}
  {\bibfnamefont {W.~N.}\ \bibnamefont {Hardy}}, \bibinfo {author}
  {\bibfnamefont {D.~A.}\ \bibnamefont {Bonn}}, \bibinfo {author}
  {\bibfnamefont {M.}~\bibnamefont {Raudsepp}}, \bibinfo {author}
  {\bibfnamefont {N.~J.~C.}\ \bibnamefont {Ingle}},\ and\ \bibinfo {author}
  {\bibfnamefont {A.}~\bibnamefont {Damascelli}},\ }\bibfield  {title}
  {\bibinfo {title} {\ce{Tl2Ba2CuO_{6+\delta}} brings spectroscopic probes deep
  into the overdoped regime of the high-${T}_{c}$ cuprates},\ }\href
  {https://doi.org/10.1088/1367-2630/9/2/028} {\bibfield  {journal} {\bibinfo
  {journal} {New J. Phys.}\ }\textbf {\bibinfo {volume} {9}},\ \bibinfo {pages}
  {28} (\bibinfo {year} {2007})}\BibitemShut {NoStop}%
\end{thebibliography}%

\end{document}